\documentclass[10pt,journal,letterpaper,compsoc]{IEEEtran}

\usepackage{tabularx}
\usepackage{multirow}
\usepackage{algorithm}
\usepackage{algorithmicx}
\usepackage{color}
\usepackage{url}
\usepackage{textcomp}
\usepackage{graphicx}
\usepackage{amssymb,amsmath}

\newtheorem{theorem}{Theorem}
\newtheorem{definition}{Definition}

\newtheorem{lemma}{Lemma}

\newtheorem{question}{Question}
\newtheorem{example}{Example}
\ifodd 112
\newcommand{\rev}[1]{{\color{blue}#1}} %revise of the text
\newcommand{\revj}[1]{{\color{red}#1}} %revise of the text
\newcommand{\revh}[1]{{\color{magenta}#1}} %revise of the text
\newcommand{\com}[1]{\textbf{\color{red} (COMMENT: #1) }} %comment of the text
\newcommand{\comg}[1]{\textbf{\color{green} (COMMENT: #1)}}
\newcommand{\response}[1]{\textbf{\color{green} (RESPONSE: #1)}}  %comment of the text
\else
\newcommand{\rev}[1]{#1}
\newcommand{\revj}[1]{#1}
\newcommand{\revh}[1]{#1}
\newcommand{\com}[1]{}
\newcommand{\comg}[1]{}
\newcommand{\response}[1]{}
\fi

\ifodd 12
\newcommand{\revxx}[1]{{\color{blue}#1}} %revise of the text
\else
\newcommand{\revxx}[1]{#1}

\fi

\newcommand{\revmm}[1]{{#1}}
\newcommand{\revww}[1]{{#1}} %revise of the text in the 3rd version

\newcommand{\revv}[1]{{#1}}
\newcommand{\comv}[1]{}

\def	\M{\mathcal{M}}
\def	\N{\mathcal{N}}
\def	\eq{\triangleq}

\def	\PT{\mathrm{PT}}
\def	\PR{\mathrm{PR}}
\def	\ST{\mathrm{ST}}
\def	\SR{\mathrm{SR}}

\def	\SNR{\kappa}

\def	\POEQ{\textsf{PU-Optimal-EQ}}
\def	\RBEQ{\textsf{PU-Robust-EQ}}

\newcommand{\footnotesc}[1]{\footnote{#1}}
\begin{document}

\title{Two-sided Matching Based Cooperative Spectrum Sharing}

\author{Lin Gao,~\IEEEmembership{Member,~IEEE,}
		Lingjie~Duan,~\IEEEmembership{Member,~IEEE,}
        and~Jianwei~Huang,~\IEEEmembership{Fellow,~IEEE}% <-this % stops a space
\IEEEcompsocitemizethanks{
\IEEEcompsocthanksitem
Lin Gao is with the School of Electronic and Information Engineering, Harbin Institute of Technology Shenzhen Graduate School, China, Email: gaolin@hitsz.edu.cn; 
Lingjie Duan is with Engineering Systems and Design Pillar, Singapore University of Technology and Design (SUTD), Singapore, Email: lingjie\_duan@sutd.edu.sg;
Jianwei Huang (corresponding author) is with the Network Communications and Economics Lab, Department of Information
Engineering, The Chinese University of Hong Kong, Email: jwhuang@ie.cuhk.edu.hk. 
\IEEEcompsocthanksitem
This work is supported by the General Research Funds (Project Number CUHK 412713) established under the University Grant Committee of the Hong Kong Special Administrative Region, China.
 }}

\IEEEcompsoctitleabstractindextext{%
\begin{abstract}
%\comv{The first several sentences can focus on a bigger picture. }
\revv{Dynamic spectrum access (DSA) can effectively improve the spectrum efficiency and alleviate the spectrum scarcity, by allowing unlicensed secondary users (SUs) to access the licensed spectrum of primary users (PUs) opportunistically.
%Providing proper \emph{economic incentives} for both PUs and SUs is essential for the success of DSA.
%One effective way to achieve this goal is the \emph{cooperative spectrum sharing}, wherein SUs relay PUs' traffic in exchange for the opportunities to access to PUs' spectrum.
%Providing proper \emph{incentives} for primary users (PUs) and secondary users (SUs) to share licensed spectrum is essential for the practical implementation of dynamic spectrum access.
%Cooperative spectrum sharing is one of the promising paradigms to achieve this goal,
Cooperative spectrum sharing is a new promising paradigm to provide necessary \emph{incentives} for both PUs and SUs in dynamic spectrum access.
The key idea is that SUs relay the traffic of PUs in exchange for the access time on the PUs' licensed spectrum.
In this paper, we formulate the cooperative spectrum sharing  between multiple PUs and multiple SUs as a \emph{two-sided market},
%and use matching theory to
%A key problem arising in such a scenario is: \emph{who (SU) will relay whose (PU) traffic, and what will be the SU's relay effort and PU's time reward?}
%We solve the problem using the {matching theory}.
%Specifically, we formulate the problem as a \emph{two-sided matching} market,
and study the market equilibrium under both complete and incomplete information.
First, we characterize the sufficient and necessary conditions for the market equilibrium.
We analytically show that there may exist multiple market equilibria, among which there is always a unique \emph{Pareto-optimal} equilibrium for PUs (called {\POEQ}),
%where \emph{every} PU achieves its maximum utility among all possible equilibria.
in which \emph{every} PU achieves a utility no worse than in any other equilibrium.
%with PUs as one side and SUs as the other side.
%A \emph{market equilibrium} is defined as a {stable matching} between PUs and SUs, where no player has the incentive to deviate.
%Then, we study which specific market {equilibrium} will actually emerge in different information scenarios.
%First, we characterize the sufficient and  necessary  conditions for equilibrium, and analytically show that there are multiple equilibria, among which there exists a unique \emph{Pareto-optimal} equilibrium for PUs, where \emph{every} PU achieves its maximum utility among all possible equilibria.
% a utility no worse than under any other equilibrium.
Then, we show that under complete information, the unique {Pareto-optimal} equilibrium  {\POEQ} can always be achieved despite the competition among PUs; whereas, under incomplete information, the {\POEQ} may not be achieved due to the mis-representations of SUs (in reporting their private information).
Regarding this, we further study the worse-case equilibrium for PUs, and characterize a \emph{Robust} equilibrium  for PUs (called {\RBEQ}),
% depending on the amount of information known by PUs,
which provides \emph{every} PU a guaranteed utility under all possible mis-representation behaviors of SUs.
%which ensure SUs' truthful information disclosure, and provide the performance lower-bounds for PUs.
%Finally, we propose ``deferred acceptance'' algorithms to realize the PU-optimal and PU-pessimistic equilibria.
%Our numerical results show the impacts of network information and market structure on the achieved equilibrium.
%Interestingly, we show that in an un-balanced market (
Numerical results show that in a typical network where the number of PUs and  SUs are   different, the performance gap between {\POEQ} and {\RBEQ} is quite small (e.g., less than 10\% in the simulations).}
%This implies that the Pareto-optimal equilibrium or the robust equilibrium is a good approximation of any equilibrium.

\vspace{-2mm}

\end{abstract}

}

% make the title area
\maketitle

\IEEEdisplaynotcompsoctitleabstractindextext

\IEEEpeerreviewmaketitle

%!TEX root = CoopSpecShar_main.tex
%SourceDoc CoopSpecShar_main.tex

\section{Introduction}\label{sec:intro}

%1. "I would add a little more discussion about the relevance and
%motivation of these models in cognitive radio problems. For example,
%how can the results influence regulation and what is the anticipated
%market behavior?" This is very very important. We need to motiviate
%more about why this approach can be useful in the near future. One key
%reason to emphasize: economic incentives for both PUs and SUs. Mention
%that such idea can be generalized to other types of PU-SU
%interactions. Also, let us fully utilize the abstract and conclusion
%to highlight the technical challenges and practical impact, besides
%the technical results.
%

%Frequency spectrum is a limited resource for wireless communications, and is becoming increasingly congested with the explosive development of wireless services and networks.
%These observations point to the major factor that leads to the inefficient use of  spectrum is the traditional spectrum allocation scheme itself, with which the spectrum allocated to licensed users cannot be utilized by unlicensed users, even if it is not used by its licensees.
%Cognitive radio based dynamic spectrum access is a novel approach to increase spectrum efficiency and alleviate spectrum scarcity. The key idea of cognitive radio is to allow  unlicensed secondary users (SUs) to access the licensed spectrum  opportunistically
% \cite{haykin2005cognitive, akyildiz2006next, zhao2007survey, buddhikot2007understanding}.

\subsection{Background and Motivation}

Wireless spectrum is becoming increasingly congested and scarce with the explosive development of wireless devices and services.
\rev{Dynamic spectrum access (DSA)} is a promising approach to increase the spectrum efficiency and alleviate the spectrum scarcity, by allowing unlicensed secondary users (SUs) to opportunistically access to the spectrum licensed to primary users (PUs) \cite{yuan2007allocating}-\cite{wu2006distributed}.
%\cite{yuan2007allocating, baocun2012, murty2011senseless, xu2010efficient, wu2006distributed}
%\revmm{Hence, it has become one of the important Information and Communication Technologies (ICT) in ``Smart City''.}
%\footnote{\revmm{Smart City \cite{app-1} is a new concept that brings together cities, industry, and citizens to improve urban life through more sustainable integrated solutions.
%The intelligent use of ICT such as DSA is an important part of a smart city \cite{app-2,app-3}.}}
To successfully implement DSA, it is important to offer necessary incentives for PUs
to open their spectrum for SUs' utilizations, \rev{and for SUs to access the PUs' spectrum despite of the potential costs  \cite{zhou2009trust}-\cite{xwang8}.}
%\revj{Without proper incentives, it can be quite difficult to implement DSS in practice. The law suit between the US National Association of Broadcasters (NAB) and Federal Communications Commission (FCC) in 2009 regarding the sharing of TV spectrum illustrates this point well \cite{lawsuit}.}

\begin{figure}[tt]
\vspace{-5mm}
\centering
 \includegraphics[width=0.45\textwidth]{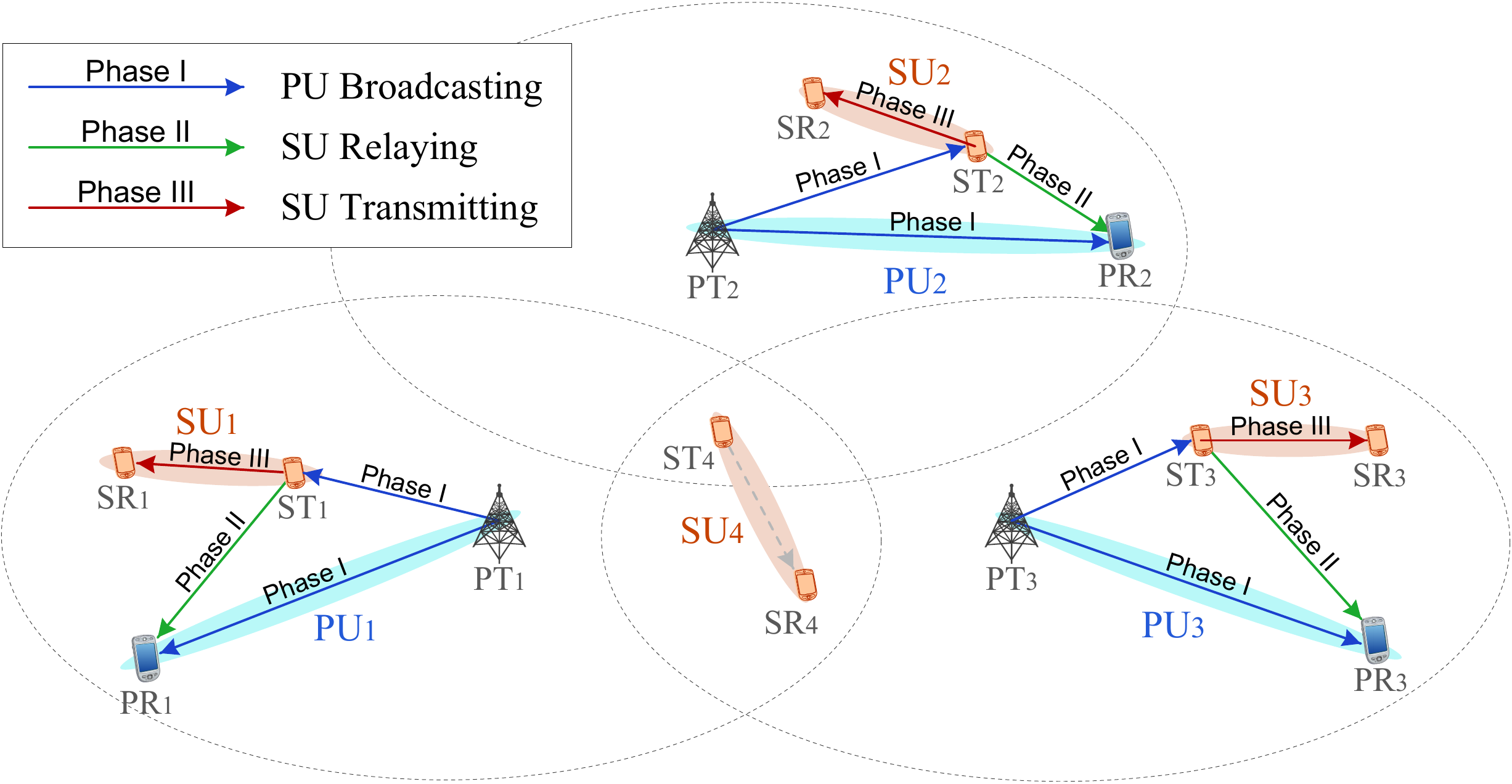}
\caption{Illustration of  cooperative spectrum sharing.
%Each SU $i\in\{1,2,3\}$ cooperates with each PU $i\in\{1,2,3\}$, and SU $i=4$ does not cooperate with any PU.
%Phase I: each PU $i$ broadcasts messages to its receiver $\PR_i$ and the cooperating SU $i$;
%Phase II: each SU $i$ forwards the received PU $i$'s signal to $\PR_i$.
%Phase III: each SU $i$ transmits its own messages on PU $i$'s spectrum.
}
\label{fig:network-illu}
\vspace{-4mm}
\end{figure}

%\cite{simeone2008spectrum}
Cooperative spectrum sharing ({also called cooperative spectrum leasing \cite{simeone2008spectrum}}) \rev{has been proposed as an effective approach} to offer crucial incentives for \revj{both} PUs and SUs in DSA \cite{simeone2008spectrum}-\cite{bayat2011cognitive}.
%\footnotesc{There are some other work on incentives issues, e.g., spectrum leasing (\cite{jayaweera2009dynamic,niyato2008market,wang2008price}). In spectrum leasing, a PU will lease its temporarily unused spectrum to SUs by charging some money. However, the success of spectrum leasing heavily relies on whether we can build up a trustworthy billing system, which is   far from implementation by now.}
The basic idea is that SUs relay the traffic of PUs in exchange for the opportunities to access the PUs' licensed spectrum.
With the cooperative spectrum sharing, PUs can increase their transmission rates through the cooperative relay of SUs, and SUs can obtain transmission opportunities on the PUs' licensed spectrum.
Thus, it will lead to a win-win situation for PUs and SUs.\footnote{\revww{Cooperative spectrum sharing can also be viewed as an enhanced cooperative relay scheme.
From the perspective of cooperative
relays (e.g., D2D relays in 5G system \cite{5G}), it can provide the necessary incentives for SUs to relay the traffic of PUs.}}
Figure~\ref{fig:network-illu} illustrates \revv{an example of cooperative spectrum sharing between 3 PUs and 4 SUs},\footnote{Here each PU $i$ is a dedicated transceiver pair \{$\PT_i$, $\PR_i$\}, and each SU $i$ is also a dedicated transceiver pair \{$\ST_i$, $\SR_i$\}.} where each SU $i $ cooperates with   PU  $i $ for $i \in\{1,2,3\}$, and SU $4$ does not cooperate with any PU.
\revmm{SUs relay traffic for intended PUs  according to a pre-defined cooperative protocol, such as amplified-and forward (AF) and decode-and-forward (DF).}
Hence, each cooperation frame is divided into 3 phases:
%\footnote{\revmm{Note that with different cooperative protocols, the cooperation frame may be different from that in Figure~\ref{fig:network-illu}. Our analysis techniques in this work are general, and can be applied to other cooperative protocols with minor modifications.}}
in Phase I, each PU $i$'s transmitter ($\PT_i$) broadcasts its messages to its receiver ($\PR_i$) as well as to the cooperating SU $i$;
in Phase II, each SU $i$'s transmitter ($\ST_i$) forwards the received PU $i$'s signal to the PU $i$'s receiver $\PR_i$;
and in Phase III, each SU $i$ transmits its own messages (from $\ST_i$ to $\SR_i$) on the PU $i$'s spectrum.

While there are some prior works considering the cooperative spectrum sharing problem \cite{simeone2008spectrum}-\cite{bayat2011cognitive}, all of these works focused on the interactions between \emph{one} PU and one or multiple SUs.
Our work advances the research in this area by analyzing the interaction between \emph{multiple PUs and multiple SUs} as shown in Figure \ref{fig:network-illu}.
This more practical scenario is significantly more challenging to analyze, as not only SUs compete with each other for the PUs' licensed spectrums, but PUs also compete with each other for the SUs' collaborations.
%The key questions arising in such a scenario are therefore: \emph{who will relay whose traffic, and how?} We answer these questions by using the \emph{matching} theory.
%We want to answer the following
An important question arising in such a multi-PU multi-SU scenario is:
\begin{itemize}
\item
\emph{Which PU cooperates with which SU, and what is the resource exchange (i.e., the PU's spectrum access time reward and the SU's relay effort) between them?}
\end{itemize}
%how they agree to exchange the (PU's) spectrum resources and the (SU's) relay efforts?}
As the traditional matching problem usually studies the binary pairing of users without considering the resource exchange details, our problem is actually an extended matching problem.
%Thus, we will study the above problem by using the matching theory.

\vspace{-3mm}
%We try to answer this question using the \emph{matching theory}.
\subsection{Solution and Contribution}

In this work, we formulate the cooperative spectrum sharing problem between multiple PUs and multiple SUs as a \emph{two-sided matching} market, with PUs as one side and SUs as the other side.
A PU is matched to an SU means that the PU cooperates with the SU under certain resource exchange agreement.
Accordingly, a \emph{matching} between PUs and SUs defines not only their collaborative relationships, but also the resource exchange between each pair of matched PU and SU.
An \emph{equilibrium} is defined as a \emph{stable} matching, from which neither PUs nor SUs have the incentive to deviate.
%We will study the market equilibrium in different information scenarios systematically.
For such a two-sided market, we want to answer the following questions:
\begin{itemize}
   \item \emph{what is the market equilibrium, and which equilibrium will emerge in different information scenarios?}
 \end{itemize}
Note that both problems are challenging due to the following reasons.
\revv{First, finding the stable matching of a two-sided matching market is well-known an NP-hard problem, and thus mathematically intractable.
Second, in some two-sided matching markets including the one in this paper, there may be an infinite number of market equilibria, and analytically characterizing these market equilibria is  challenging.
Third, different information scenarios (depending on how much information that PUs know) may lead to different market equilibria.
Characterizing the equilibrium in different information scenarios is also a challenging problem.}

We will study the market equilibrium in two different information scenarios systematically: \emph{complete} and \emph{incomplete} information.
In the former case, each PU knows the whole network information, while in the latter case, each PU knows only its local network information (see Section \ref{sec:model}.F for details).
To the best of our knowledge, this is the first paper that systematically studies cooperative spectrum sharing between multiple PUs and multiple SUs by using the matching theory.
The main results and contributions of this paper are summarized as follows.~~~~
\revv{
\begin{itemize}

\item \textbf{\emph{Two-sided Market Model and Solution Technique:}}
We formulate the cooperative spectrum sharing between multiple PUs and multiple SUs as a two-sided matching market, and comprehensively study the market equilibrium (stable matching) in both complete and incomplete information scenarios.~~~~~~~~~~~
    %We prove the  necessary and sufficient conditions for equilibrium and analyze the equilibria under different information scenarios systematically.

%\item \emph{Multiple information scenarios:}
%    We consider 3 different information scenarios: complete information, weakly incomplete information, and strongly incomplete information, and study the corresponding equilibria comprehensively. % For the first case, we mainly focus on the collective interactions between users. For the latter two cases, we consider the incentive for users truthfully representing their preferences.

\item \textbf{\emph{Equilibrium Analysis:}}
We characterize the sufficient and necessary conditions for market equilibrium, and show that there may be an infinite number of market equilibria.
We further show that these is always a unique Pareto-optimal equilibrium for PUs ({\POEQ}) , in which \emph{every} PU achieves a utility no worse than that in any other equilibrium.~~~
%\footnote{\revv{Notice that in a symmetric network where all PUs are identical or all SUs are identical, there may be multiple symmetric Pareto-optimal equilibria, but every PU achieves the same maximum utility under these Pareto-optimal equilibria comparing to other equilibria.}\comm{may be not a good idea to present here and can be moved to some section later}}

\item \textbf{\emph{Complete and Incomplete Information:}}
We show that under complete information, the unique Pareto-optimal equilibrium for PUs ({\POEQ})  can always be achieved despite  of the competition among PUs.
%We further introduce a Generalized Deferred Acceptance (G-DAC) algorithm, which converges to the {\POEQ} in polynomial time.
%\item \textbf{\emph{Incomplete Information:}}
Under incomplete information, however, the {\POEQ} may \emph{not} be achieved due to the mis-representations of SUs.
%Moreover, it is difficult to characterize all possible mis-representation behaviors of SUs.
To this end, we characterize a \emph{Robust} equilibrium  for PUs ({\RBEQ}), which provides \emph{every} PU a guaranteed utility under all possible mis-representation behaviors of SUs.
%We further introduce a Generalized Reverse Deferred Acceptance (G-RDAC) algorithm, which converges to the {\RBEQ} in polynomial time.

\item \textbf{\emph{Performance Evaluation:}}
Numerical studies show that each PU's utility increases (or decreases) with the number of SUs (or PUs) in both {\POEQ} and {\RBEQ}.
\revv{In many practical network scenarios where the numbers of PUs and SUs are often different, the performance gap between {\POEQ} and {\RBEQ} is quite  small (e.g., less than 10\% in the  simulations).
%This implies that the impact of market structure dominates compared to the loss of network information.
This implies that the equilibrium performance depends more on the market structure than on the information scenario.}~~~~
\end{itemize}}

%
%\revyy{Our work is instructive in designing large networks where two types of players (e.g., PUs and SUs) interact with each other independently. The interaction between players is not restricted within the cooperative spectrum sharing mentioned in this work, but can be quite general (e.g., trading, exchanging, etc.). The equilibrium in this work provides a well prediction for the \emph{network stable state} resulting from the collectively interaction between players. Furthermore, the deferred acceptance algorithms provide methods to converge to the stable state in a decentralized manner. It is worthy noting that the prediction of a large network's stable state is nontrivial in practice, and can provide instruction in different objectives, e.g., network optimizating. }

The rest of this paper is organized as follows.
In Section~\ref{sec:liter}, we review the related literature.
In Section~\ref{sec:model}, we present the system model.
In Section~\ref{sec:market-formulation}, we provide the two-sided market formulation.
In Section~\ref{sec:toy}, we use a simplified example to illustrate the equilibrium.
In Section~\ref{sec:main}, we study the equilibria of a general model.
In Section~\ref{sec:main2}, we show which equilibria will emerge in different information scenarios.
We provide numerical results in Section~\ref{sec:simu}, and finally conclude in Section~\ref{sec:con}.

%!TEX root = CoopSpecShar_main.tex
%SourceDoc CoopSpecShar_main.tex

\section{Literature Review}\label{sec:liter}

%
%\subsubsection*{A. Cooperative Spectrum Sharing}
%
%\revxx{Dynamic spectrum sharing is a promising approach to increase spectrum efficiency and alleviate spectrum scarcity (e.g., \cite{yuan2007allocating, baocun2012, murty2011senseless, xu2010efficient, wu2006distributed}).
%Providing proper economic incentives for both PUs and SUs is essential for
%the success of dynamic spectrum sharing, but related studies only emerge recently (e.g.,  \cite{zhou2009trust, zhou2008traffic, xwang8}).
%Cooperative spectrum
%sharing is one effective way to achieve this goal.}

%There are several comprehensive surveys on DSS \cite{haykin2005cognitive, akyildiz2006next}. While many results focused on the technical aspect of DSS, we consider the issue of economic incentives that are essential for the success of DSS.

\subsubsection*{A. Cooperative Spectrum Sharing}
Most prior work on cooperative spectrum sharing focused on the interactions between \emph{one} PU and one or {multiple SUs} \cite{simeone2008spectrum}-\cite{bayat2011cognitive}. Some assumed that the PU has complete information of SUs (e.g., SUs' relay channel gains and sensitivities to power consumption) \cite{simeone2008spectrum}-\cite{wang2010cooperative}, while others considered that the PU has limited information about SUs \cite{Yan2011,duan2011contract}.
%In \cite{simeone2008spectrum, zhang2009stackelberg, wang2010cooperative, han2008cooperative}, researchers consider the complete information scenario where the PU has full knowledge about SUs' wireless characteristics (e.g., relay channel gains and evaluations of power consumptions).
%In \cite{Yan2011,duan2011contract}, researchers consider the incomplete information scenario where the PU has limited knowledge about SUs' information.
%Specifically, in \cite{Yan2011}, Yan \emph{et al.} study the interaction between one PU and one SU using dynamic Bargaining game.
%In \cite{duan2011contract}, Duan \emph{et al.} study the interaction between one PU and multiple SUs using contract theory.
%However, this assumption may not hold in practical relay networks. Some SUs' own transmissions may vary from time to time, depending on their temporary available spectrum and locations (if mobile), thus it is hard for PUs to monitor their wireless characteristics.
%Even some work also consider incomplete information, they do not study the interactions between \emph{multiple PUs} and \emph{multiple SUs} (e.g., \cite{Yan2011,duan2011contract}). Yan et al. in \cite{Yan2011} proposes a dynamic Bargaining game to study the interaction between one PU and one SU over time. Duan et al. in \cite{duan2011contract} uses contract theory to study the interaction between one PU and multiple types of SUs.
%These works do not consider the interactions between \emph{multiple PUs} and \emph{multiple SUs}, and besides, may not be easily extended to the general model with multiple PUs.
These works used either Stackelberg game or contract theory to model and analyze the problem. However, both approaches are difficult to be extended to the scenario of multiple PUs and multiple SUs.
A closely related paper that considers the interactions between multiple PUs and multiple SUs is \cite{bayat2011cognitive}, where authors proposed a distributed algorithm to reach a stable matching that is weak Pareto optimal.
Our work generalizes the result of \cite{bayat2011cognitive} in the following way:
(i) we analytically show that there are multiple (possibly infinite) stable matchings including the Pareto optimal one studied in \cite{bayat2011cognitive},
(ii) we characterize the necessary  and sufficient conditions for all possible stable  matchings,
 and
 (iii)  we propose distributed algorithms converging to different stable matchings, given different information available to PUs.

\subsubsection*{B. Two-sided Matching Market}
In economics, two-sided matching market is an effective framework and widely-used for studying the interactions between two disjoint player sets in a two-sided market setting.
\revmm{In such markets, the matching theory can systematically capture not only the cooperative interactions between users in different sides, but also the competitive interactions between users on the same side.}
%\footnote{\revmm{Note that the matching theory is a widely-used analytical tool for two-sided markets with no effective central
%controller, where each user in one side can freely choose to cooperate with any other user in the
%other side. In such markets, the matching theory can effectively capture not only the cooperative
%interactions between users in different sides, but also the competitive interactions between users in
%the same side. Moreover, the outcome resulting from the matching theory is the stable matching
%(i.e., equilibrium), which disclose what state the system will naturally evolve to in reality.}}
\rev{Most early results in this area focused on the matching under complete information, without considering the incentive issues under incomplete information.}
%, and it has been extensively investigated in the literature of economics.
The first basic two-sided market models were proposed by Gale and Shapley \cite{gale1962college}.
Shapley and Shubik \cite{shapley1971assignment} and Thompson \cite{Thompson1980Core} studied the more general models with additive and transferable utilities.
Crawford and Knoer \cite{crawford1981job, kelso1982job} studied the two-sided labor models with nontransferable utilities.
Some later results \cite{roth1984misrepresentation, kojima2009incentives} studied \rev{the incentive issue} under \revj{incomplete} information.
%\com{I have removed the literature surveys.}
%Moreover, Roth provided several comprehensive surveys on the two-sided matching market models  in \cite{roth1992two, roth2008deferred}.
%The PUs compete with each other by offering transmission opportunities (time) to SUs to attract the SUs' relaying effort (power), and the SUs compete with each other by offering relaying efforts to PUs to compete for the transmission opportunities offered by the PUs.
Our work differs from the above works:
we consider not only the binary matching decision, but also the detailed resource exchange between each pair of matched users.
%(i) we consider a two-sided market with \emph{non-additive} and transferrable utilities, \revj{which renders the analysis more challenging};
%%\com{is this new in the economics literature? If so, we should mention it.}
%(ii) we study the equilibria under different information scenarios systematically.

\revmm{It is notable that our analysis techniques and engineering insights are applicable to other wireless network problems that can be  modeled as a two-sided market.
The equilibrium obtained in this paper provides a quite general characterization of the network stable outcome in a wide range of large complicated networks.}
%Moreover, the proposed algorithms enable distributed convergence to the stable outcome of this type of networks.}

%!TEX root = CoopSpecShar_main.tex
%SourceDoc CoopSpecShar_main.tex

\section{System Model}\label{sec:model}

We consider a DSA network with a set $\M \eq\{1,...,M\}$ of  PUs and a set $\N\eq\{1,...,N\}$ of SUs. Each PU $m\in\M$ (or SU $n\in\N$) is a dedicated transceiver pair $\{\PT_m,\PR_m\}$ (or $\{\ST_n,\SR_n\}$) as illustrated in Figure \ref{fig:network-illu}.
%each PU $m\in\M $ (or SU $n\in\N $) is a dedicated transceiver pair $\{\PT _m$, $\PR _m\}$ (or $\{\ST _n$, $\SR _n\}$).
A PU has the exclusive usage right of a licensed frequency band (with a normalized bandwidth). \revmm{We assume that the frequency bands of different PUs are \emph{non-overlapping}, and thus there is no interference among PUs' transmissions.}
\revww{As in many existing literature \cite{simeone2008spectrum}-\cite{bayat2011cognitive}, we further assume that there exists a common control channel for the communications and interactions between PUs and SUs.}

%\footnotesc{\revyy{When PUs transmit on a common band in the coordinated manner (e.g., TDMA and OFDMA), our analysis is directly applicable. However, when PUs transmit in the uncoordinated manner (e.g., CSMA) or even concurrently (e.g., CDMA), the cooperation between a PU and an SU may cause additional interference to other PUs' or SUs' transmission. This essentially leads to a \emph{partition problem} (where the gain of each group depends on the strategies of all members), rather than a \emph{matching problem} (where the gain of each group (a matching pair) depends on the strategies of its own members only).  We leave the partition modeling in our future work.}}

On one hand, PUs may suffer from low transmission rates or high outage probabilities due to the poor channel conditions between their transmitters and receivers caused by, for example, the long-distance attenuation, shadowing, and fading.
%the shadow fading cased by large distances between their respective transmitters and receivers, the fast fading caused by multiple reflections coinciding on the receivers' antennas and cancelling, the severe interference from outside environment, and etc.
%Thus, each PU is willing to select SUs to cooperatively relay its traffic, as long as such selection can increase the PU's own transmission rate.
%Besides, the SUs are eager for spectrum for their own transmission demands.
%Thus, each SU expects to get certain dedicated access time on some licensed frequency bands.
\rev{Thus, PUs want to employ SUs as relays to improve their  transmission rates.} %\com{should I say ``one SU'' or ``one or more SUs''?}
%An SU {is unlicensed and} cannot use any spectrum without the explicit permission of a PU, and thus has the incentive to relay PUs' traffic in order to get spectrum resource in return.
On the other hand, SUs need spectrum resources for their own transmissions, but cannot access the PUs' licensed spectrum without PUs' permissions.
Thus, we expect a win-win situation where SUs relay PUs' traffic in order to get free access time on the PUs' spectrums.

\rev{To facilitate the practical implementation, we assume that one PU can choose \emph{at most one} SU to relay its traffic at a particular time (but can change the relaying SU at different time).}
This is motivated by \revj{existing results} that choosing \revj{the most appropriate} relay is \revh{usually} \revj{sufficient to achieve the optimal} (or close-to-optimal) \revj{performance} \cite{yi2010cooperative}.
%\footnotesc{For many cooperative protocols (e.g., that in \cite{duan2011contract}), it is \emph{optimal} for the PU to select one (the most suitable) SU as relay.}
Moreover, this can be implemented more easily in practice, \revj{as there is no need to consider the coordination among multiple relays}.
\rev{We further assume that} each SU can serve \emph{at most one} PU at a particular time (but can change the serving PU at different time).
%This is because radio frequency (RF) devices usually operate one frequency band.
Therefore, a key question arising in such a scenario is:
\textbf{\emph{who will relay whose traffic, and how?}}

%\footnotesc{Our results can be extended to the general case where one SU \revh{(equipped with multiple antennas)} can serve as multiple PUs' relays simultaneously. \revj{Mathematically, we can split such an SU} into multiple virtual SUs. \revh{Each SU's utility is simply the summation of all associated virtual SUs' utilities, since both transmission rate and power consumption are additive.}}

%We assume an SU can only access one frequency band at the same time due to, for example, radio hardware restrictions, and thus can only server as \emph{one} PU's relay.
%We further suppose a cooperative relaying protocol where a PU can only select \emph{one} SU as   relay.
%{The motivation is multi-fold: First, for many cooperative relaying protocols, it is indeed optimal for the PU to select one (the most suitable) SU as relay (e.g., \cite{duan2011contract}); Second, for those models that selecting multiple relays is possibly optimal, the gap between the optimal performance and the performance with one properly chosen relay is very small (see \cite{zhao2007improving});
%Third, this is also motivated by the practical implementation of one relay network \cite{yi2010cooperative}.}

\begin{figure}[tt]
\vspace{-3mm}
\centering
\includegraphics[width=0.49\textwidth]{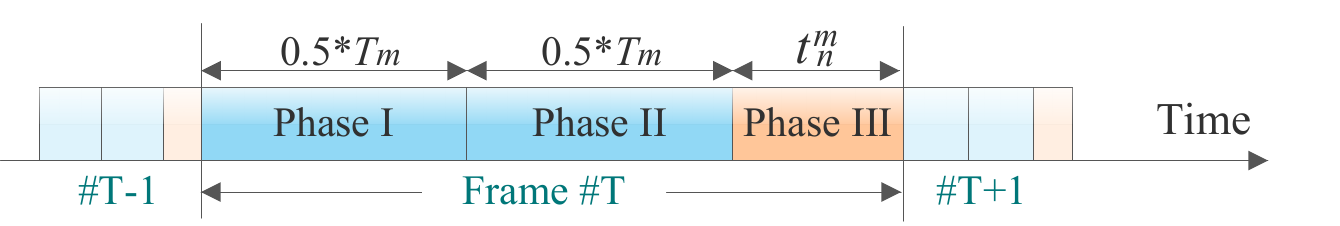}
\caption{Frame structure of cooperative spectrum sharing.}
\label{fig:network-frame}
\vspace{-4mm}
\end{figure}

\subsection{Cooperative Spectrum Sharing Protocol}

%\revv{Suppose the cooperative relationships between PUs and SUs are given.
We first consider the cooperative spectrum sharing protocol between a particular PU $m$ and SU $n$ (assuming SU $n$ cooperates with PU $m$).
%The PU's transmission is organized into frames
As shown in Figure~\ref{fig:network-illu},
%the cooperative spectrum sharing between PU $m$ (or $\{\PT_m, \PR_m\}$) and SU $n$ (or $\{\ST_n, \SR_n\}$)
each cooperation period (frame) includes 3 phases:
Phases~I and II for the cooperative communication between PU $m$ and SU $n$ (each with a fixed period of $T_m/2$),\footnotesc{The total cooperative transmission period $T_m$ is \rev{determined} by the physical layer specifications. \revj{Different PUs may have different $T_{m}$.}}
and Phase~III for SU $n$'s own transmission (with a period of $t_n^m$). %\footnotesc{The period $T_m$ is a constant predefined by the PU $m$'s mac layer and physical layer specifications.}
%For many cooperative communication protocols (e.g., {Amplified-and-Forward} and {Decode-and-Forward}), the whole cooperative transmission period $T_m$ is further divided into two slots with the same size (i.e., $T_m/2$): one for the PU's broadcasting and the other for the SU's relaying.
Thus, the total length of one cooperation frame is $T_m+t_n^m$.
For clarity, we illustrate the structure of such a 3-phase transmission frame in
Figure \ref{fig:network-frame}.
Obviously, the cooperative communication protocol used in Phases I and II plays an important role in the cooperative spectrum sharing.
In this paper, we adopt the widely-used \emph{Amplified-and-Forward} (AF) protocol \cite{laneman2004cooperative} in Phases I and II.\footnote{Note that our analysis can be easily \rev{applied} to other cooperative protocols (e.g., Decode-and-Forward  and Compress-and-Forward) with minor modifications on the utility functions of PUs and SUs.}
For the analytical convenience, we assume the \emph{AWGN channel} in the physical links.
\revj{Note that the channel model can be easily extended to} fast fading channels (e.g., Rayleigh channel),
by considering the average transmission rate or average outage probability.
% taking into account the channel's statistic characteristics such as the mean or variance of channel gain.
%since the PUs' and SUs' \revh{utility} functions are directly applicable to the fast fading channels.
We further assume that each pair of transmitter and receiver know the channel gain between them through, for example, measuring the received signal strength.
%\footnote{\revmm{Difference between our model and traditional cooperative communication schemes are as follows.
%Traditional cooperative communication
%schemes usually assume that the relays (corresponding to the SUs in our model) are \textit{altruistic} and \textit{compliant}, hence researchers usually focus on the design and optimization of the cooperative transmission protocols for the primary source-destination pairs (corresponding to the PUs in our model), without considering incentives and QoS issues for the relays.
%In our work, however, we assume that SUs are \textit{selfish} and \textit{strategic}, which is reasonable in the context of dynamic spectrum sharing as   SUs do not belong to the same system as    PUs.
%Our cooperative spectrum sharing scheme consists of
%a cooperative communication scheme for PUs (in Phase I and Phase II) and a secondary
%transmission stage as the incentive for SUs (in Phase III).}}~~~~~~~~

%Let $G_m$ denote the channel gain of the PU $m$'s direct channel, $G_{n(m)}$ denote the channel gain of the SU $n$'s direct channel on the PU $m$'s frequency band, and $G_{m,n}$

\revh{For convenience, we list the key notations in Table \ref{tbl:1}, where $m$ is the PU index and $n$ is the SU index. The meaning of each notation will be explained later.}

\subsection{PU and SU Modeling}\label{subsec:PU}

We now define the PU's and SU's utilities achieved in the cooperative spectrum sharing.
%Consider the cooperative spectrum sharing between PU $m$ and  SU $n$ (assuming SU $n$ collaborates with PU $m$).
Suppose SU $n$ cooperates with PU $m$ in the following way: SU $n$ relays the PU $m$'s signal with a power $p_n^m$, and PU $m$ rewards a dedicated access time $t_n^m$ on its spectrum to SU $n$.~~~~~~~

\begin{table}[t]
\small
    \caption{Key Notations}\label{tbl:1}
    \hspace{-5mm}
\begin{tabular}{m{0.028\textwidth}m{0.45\textwidth}}
\hline
\hline
 $T_m$ & \revj{ PU $m$'s cooperative communication time (Phases I and II);}\\
 $C_n$ & \revj{SU $n$'s sensitivity of unit power consumption;}\\
%\hline
% ~ $R_m$ & The theoretical capacity of PU $m$'s direct channel;\\
% ~ $R_c^m$ & The theoretical capacity of PU $m$ with the cooperation of SU;\\
%  $R_{n(m)}$ & The theoretical capacity of SU $n$'s own channel on PU $m$' band;\\
\hline
  $G_m$ & PU $m$'s direct channel ($\PT _m$ to $\PR _m$) gain;\\
$G_{n(m)}$ & SU $n$'s direct channel ($\ST _n$ to $\SR _n$) gain
%(i.e.,   $\ST _n$ to $\SR _n$)
on \revj{PU $m$'s band;}\\
$G_{m, n}$ &   1st relay channel ($\PT _m$ to $\ST _n$) gain of PU $m$ and SU $n$;\\
% (i.e., the 1st half of the relay channel);\\
  $G_{n, m}$ &   2nd relay channel ($\ST _n$ to $\PR _m$) gain of PU $m$ and SU $n$;\\
  %(i.e., the 2nd half of the relay channel);\\
\hline
 $p_n^m$ & The relay power of SU $n$ for PU $m$'s traffic;\\
 $t_n^m$ & The dedicated access time on PU $m$'s   band for SU $n$;\\
 $\Pi_{n}^m$ & The utility of PU $m$ when cooperating with SU $n$;\\
 $\Delta_{n}^m$ & The utility of SU $n$ when cooperating with PU $m$;\\
\hline
\hline
\end{tabular}
\vspace{-4mm}
\end{table}

\rev{With the} AF protocol, the PU $m$'s transmitter $\PT _m$ broadcasts the data (with unit power) in Phase I, and SU $n$ normalizes the received signal $r_{n,m}$ by a factor ${\sqrt{p_n^m}}/{| {r}_{n,m}|}$ and forwards it to $\PR _m$ in Phase II.
The receiver $\PR _m$ combines the received signals in   \revh{Phases I and II} with a maximal ratio combining.
%Let $\texttt{x}_m$ denote the $\PT _m$'s transmitting signal with unit power in Phase I.
%Each SU $n\in \N _m$ receives signal
%\begin{equation}\label{eq:PU_n_m}
%\texttt{r}_{n,m} = \texttt{x}_m \cdot G_{m,n} + z_n,
%\end{equation}
%where $G_{m,n} $ is the channel gain between $\PT _m$ and $\ST _n$, and
%$z_n \in \mathcal{CN}(0,\sigma^2)$ is the additive white noise.
%
%By AF protocol, SU $n$ normalizes the signal by factor $\frac{\sqrt{p_n^m}}{|\texttt{r}_{n,m}|}$ and forwards the normalized signal in Phase II:
%\begin{equation}\label{eq:PU_n_m}\textstyle
%\texttt{s}_{n,m} = \texttt{r}_{n,m} \cdot \frac{\sqrt{p_n^m}}{|\texttt{r}_{n,m}|} = (\texttt{x}_m \cdot G_{m,n}  + z_n)\cdot \sqrt{\frac{p_n^m} {G^2_{m,n} + \sigma^2}} .
%\end{equation}
%
%The PU $m$'s receiver $\PR _m$ receives the integrated signal from all involved SUs in Phase II:
%\begin{equation}\label{eq:PU_n_m}
%\begin{aligned}
%\texttt{y}_m & \textstyle
% = \sum_{n\in \N _m} \texttt{s}_{n,m} \cdot G_{n,m} + z_r \\
% & \textstyle = \texttt{x}_m    \sum_{n\in \N _m} A_{m,n} + \sum_{n\in \N _m}z_n   B_{m,n}   + z_r,
%\end{aligned}
%\end{equation}
%where  $A_{m,n} \triangleq \sqrt{\frac{p_n^mG^2_{m ,n}G^2_{n,m}} {G^2_{m,n} + \sigma^2}}  $, $B_{m,n} \triangleq \sqrt{\frac{p_n^m G^2_{n,m}} {G^2_{m,n} + \sigma^2}} $ and $z_r \in \mathcal{CN}(0,\sigma^2)$ is the additive white noise.
%
%Thus, the signal-to-noise ratio of the relay channel is
%\begin{equation}\label{eq:PU_n_m}\textstyle
%\SNR_{\N _m}^m = \frac{(\sum_{n\in \N _m} A_{m,n})^2}{\sigma^2 \cdot \sum_{n\in \N _m} (B_{m,n})^2  + \sigma^2}.
%\end{equation}
Essentially, we can view the transmission during Phases I and II as a \revj{single}-input, two-output complex
Gaussian noise channel.
% with different noise levels in the outputs.
Accordingly, the transmission rate %maximal average mutual information (i.e., the capacity)
during Phases I and II
achieved from i.i.d.~complex Gaussian inputs is given by the Shannon-Hartley theorem \cite{laneman2004cooperative}:
\begin{equation}\label{eq:PU_n_m}
\textstyle
R_n^m = \frac{1}{2}\cdot \log(1 + \SNR_d^m +  \SNR_{n}^m),
\end{equation}
where $\SNR_d^m$ is the signal-to-noise ratio (SNR) on the PU $m$'s direct channel ($\PT _m$ to $\PR_m$), and $\SNR_{n}^m$ is the SNR on the relay channel of SU $n$.
For convenience, we normalize the PU's direct transmission power into 1. Then, we have
$\SNR_d^m = \frac{G_m^2}{\sigma^2}$,
where $G_m$ is the channel gain of the PU $m$'s direct channel and
$\sigma^2$ is the \revj{noise power},
%level of the additive white noise,
and
\begin{equation}\label{eq:PU_n_m}
\textstyle
\SNR_{n}^m = \frac{p_n^m G^2_{m ,n}G^2_{n,m}}{ p_n^m G^2_{n,m} + G^2_{m,n} +  \sigma^2}\cdot \frac{1}{ \sigma^2},
\end{equation}
where $G_{m,n}$ is the channel gain of the channel between $\PT _m$ and $\ST _n$ (called the 1st relay channel), and $G_{n,m}$ is the channel gain of the channel between $\ST _n$ and $\PR _m$ (called the 2nd relay channel). Obviously,
the SNR $\SNR_{n}^m$ increases with the SU's relay power $p_n^m$.

Since the cooperative period occupies $\frac{T_m}{T_m+  t_n^m}$ fraction \revh{of the total frame}, PU $m$'s \emph{effective} transmission~rate~during the entire frame (when cooperating with SU $n$) is
\begin{equation}\label{eq:PU_n_m}\textstyle
   \widetilde{R}_n^m =\frac{T_m   }{  T_m+ t_n^m }\cdot \frac{1}{2} \cdot \log(1 + \SNR_d^m +  \SNR_{n}^m) .
\end{equation}
We define  \textbf{PU $m$'s utility} as the \rev{\emph{increase}} of its transmission rate when cooperating with an SU $n$, i.e.,
%by adopting the cooperative relay, i.e.,
\begin{equation}\label{eq:PU_ut}
\begin{aligned}\textstyle
\Pi_{n}^m  \triangleq  \widetilde{R}_n^m - R_d^m= \frac{T_m\cdot \log(1 + \SNR_d^m + \SNR_{n}^m)}{2\cdot(T_m+ t_n^m)} - R_d^m,
\end{aligned}
\end{equation}
where $R_d^m =  \log(1 + \SNR_d^m ) = \log(1 + {G_m^2}/{\sigma^2} )$ is \rev{PU $m$'s  direct transmission rate without any relay.}

%\subsubsection*{C. The SU Model}\label{subsec:PU}

Similarly, the SU $n$'s transmission rate in Phase III on the PU $m$'s frequency band is given by \cite{laneman2004cooperative}:
\begin{equation}\label{eq:PU_n_m}\textstyle
R_{n(m)} \triangleq  \log\big(1 + \frac{G_{n(m)}^2}{\sigma^2}\big),
\end{equation}
where $G_{n(m)}$ is the SU $n$'s direct channel ($\ST _n$ to $\SR _n$) gain on the PU $m$'s frequency band.
%We consider the frequency selective fading, that is,
Note that each SU $n$ may have different direct channel gains $G_{n(m)}$ on different PU $m$'s band, due to the frequency selective fading.
Moreover, there is no interference between~SUs'~transmissions as they operate on non-overlapping bands.

Since the SU $n$'s transmission occupies $\frac{t_n^m}{T_m+  t_n^m}$ fraction of each frame, its  equivalent  transmission rate  during the entire frame (when cooperating with PU $m$) is
\begin{equation}\label{eq:PU_n_m}\textstyle
\widetilde{R}_{n(m)} = \frac{t_n^m}{T_m+  t_n^m} \cdot \log\big(1 + \frac{G_{n(m)}^2}{\sigma^2}\big).
\end{equation}
The SU $n$'s total energy consumption during the entire frame (when cooperating with PU $m$)  is
\begin{equation}\label{eq:PU_n_m}\textstyle
S_{n(m)} \triangleq  \frac{T_m}{2}\cdot p_n^m + t_n^m \cdot 1,
\end{equation}
where \revj{the first term is due to relaying   PU $m$'s traffic, and the second term is due to   SU $n$'s own traffic (with a normalized unit transmission power).}

We define the \textbf{SU $n$'s utility} (when cooperating with PU $m$) as the difference between the achieved transmission rate and the incurred power cost, \revh{denoted by}
\begin{equation}\label{eq:SU_ut}\textstyle
\Delta_{n}^m \triangleq  \widetilde{R}_{n(m)}- \frac{  C_n   S_{n(m)}}{T_m+   t_n^m} =\frac{ t_n^m  (R_{n(m)}-C_n) - p_n^m\frac{ C_n  T_m }{2} }{T_m+   t_n^m} ,
\end{equation}
where  $\frac{ S_{n(m)}}{T_m+   t_n^m}$ is the average transmission power,  and $C_n$ is its sensitivity for unit power consumption.
%\footnotesc{\rev{Intuitively, we can view $C_{n}$ as the equivalent data rate per unit power \revh{(that can be potentially achieved in the future)}. Then (\ref{eq:SU_ut}) is SU's data rate increase \revj{(i.e., current   rate gain minus future rate loss}.}}

%Note the term $ R_{n(m)}-C_n $ is the SU $n$'s utility from its own traffic, i.e., not considering its power consumption for PU $m$'s traffic.
%Without loss of generality, we focus on the cooperations between those PUs $m$ and SUs $n$ with $ R_{n(m)}-C_n  \geq 0$.
%For those PUs $m$ and SUs $n$ with $R_{n(m)}-C_n < 0$, we can simply ignore their cooperations from the system, since a rational SU $n$ will never accept any time from and therefore offer power to PU $m$.

\section{Two-Sided Market Formulation}
\label{sec:market-formulation}

%Each PU wants to maximize its utility by \revj{selecting} the most ``diligent'' SU as its relay, and each SU wants to maximize its utility by helping the most ``generous'' PU.
%, and as compensation, helping the band licensee (PU) with certain effort according to predefined cooperative relaying protocol.

Before the detailed modeling of the spectrum sharing market, we first provide the definition of a basic two-sided matching market (also called \emph{two-sided market}).

\begin{definition}[Two-sided Matching Market] \label{def:two-sided}
A two-sided matching market is a market consisting of two disjoint sets of users, where an user on one side can be matched with only one user on the other side.
%\footnotesc{The term ``matched'' actually means one-to-one bidirectional mapping between two disjoint sets.}
\end{definition}

A {one-to-one (binary) matching} (or just  \emph{matching}) in a two-sided market is defined as follows.

%We are interested in the one-to-one matching defined as follows.

%We will formulate the interacting between PUs and SUs as a \emph{two-sided matching market with {transferrable utilities}}. \rev{This means that} the utilities of a pair of matched PU and SU \rev{depend on the time (in Phase III) and transmission power (in Phase II)}, and \rev{a utility increase of one side leads to a utility decrease of the other side (not necessarily the same amount).} \revj{We are interested in the one-to-one matching defined as follows.}

\begin{definition}[Matching]\label{def:matching}
 A one-to-one matching between two disjoint sets $\M $ and $\N $ can be represented by a one-to-one correspondence $\mu(\cdot)$,
%if and only if $m \in \M $ is mapped to $n\in \N $ (i.e., $\mu(m)=n$) means  $n$ is   mapped to $m$ (i.e., $\mu(n)=m$), and vice versa.
%between $\M $ and $\N $,
 where $m \in \M $ is mapped to $n\in \N $ (i.e., $\mu(m)=n$) if and only if $n$ is also mapped to $m$ (i.e., $\mu(n)=m$).
\end{definition}
%\revj{For simplicity, we will simply use ``matching'' to refer to ``one-to-one matching'' later in the paper.}
%\com{See if this revision makes sense.}\response{It seems OK.}
%\com{I revised the definition of matching. See it makes sense. Also, this definition suggests that it is possible to have a matching that is NOT a one-to-one correspondence?}\response{I suggest to use the previous definition, since in general, the terminology ``matching'' denotes a one-to-one correspondence impliedly. The more important property of matching is the bidirectional mapping.}\com{Lingjie: I think we may need to change it (e.g., a matrix M*N). If not, we need to highlight it is not one-to-one mapping.}

For notational convenience, we will also write $\mu(m) $ as $\mu_m$ and $\mu(n)$ as $\mu_n$ when there is no confusion caused.

\subsection{Two-sided Market Modeling}\label{sec:model:two-sided}

In our model, each PU has different preferences over SUs depending on the locations of SUs,
and wants to select the most efficient SU as relay;
each SU also has different preferences over PUs depending on its channel gains on PUs' bands,
and want to obtain the dedicated spectrum access time from the most ``generous'' PU.
Thus,
\revj{we can formulate this model as} a \emph{two-sided market}, with PUs on one side and SUs on the other side.
%\footnotesc{A more strict discussion for such a two-sided matching formulation is provided in Appendix \ref{sec:app}.}
A PU is matched to an SU (or equivalently, an SU is matched to a PU) means \rev{that} the PU cooperates with the SU, that is,
the SU relays traffic for the PU, and the PU rewards dedicated spectrum access time to the SU.
%In this sense, a \emph{matching} specifies
Clearly, in this two-sided market, we need to consider~not~only~the binary matching between PUs and SUs (given in Definition \ref{def:matching}), but also the detailed resource exchange between each pair of matched PU and SU.\footnote{\revv{Since each binary matching is associated with a set of resource exchanges (each for a pair of matched users), we will use the terminology ``matching'' to denote both (i) the binary matching between PUs and SUs, and (ii) the resource exchange for each pair of matched users.}}
%Since an SU can only help one PU, the PUs who prefer the same SU have to compete with each other for the preferred SU.
%Since each PU can only select one SU, the SUs who prefer the same PU have to compete with each other for the preferred PU.
%
%Note that an SU can help at most one PU and a PU can select at most one SU, the question posed by the interacting between multiple PUs and SUs is therefore the following: \emph{Given each PU's preferences on SUs and each SU's preferences on PUs, what kind outcome will result from their collective interactions?}

%\revxx{Since an SU can help at most one PU and a PU can employ at most one SU, the question posed by the interacting between multiple PUs and SUs is therefore the following: \emph{Given each PU's preferences on SUs and each SU's preferences on PUs, what kind outcome will result from their collective interactions}? This is essentially a \emph{two-sided matching} between PUs and SUs (see Definition \ref{def:two-sided}).}
%\revj{We can formulate this problem as} a \emph{two-sided matching market} (see Definition \ref{def:two-sided}) between PUs and SUs.
%%\footnotesc{A more strict discussion for such a two-sided matching formulation is provided in Appendix \ref{sec:app}.}
%A PU is matched to an SU (or equivalently, an SU is matched to a PU) means \rev{that} the PU selects the SU as its relay and rewards time to the SU.

%We next provide the formal definitions for ``matching'' and ``two-sided matching'' in below for clarity.

More specifically, for each pair of matched PU and SU, the PU's and SU's utilities depend not only on the SU's relay power (in Phase II), but also on the PU's spectrum access time reward (in Phase III).
In addition, an increase in one user's utility will lead to a (not necessarily the same amount of) decrease in the other's utility in the matching pair.
\revxx{Therefore, we face a two-sided market with \emph{transferrable utility}, where each user's utility (or preference) in a given matching pair is not fixed, but depends on the resource exchange with the matched user (i.e., the access time reward and relay power).}
%Two-sided matching market is an effective model to study the collective interacting between the agents in two disjoints sets.
%In other word, each user's preference (on any possible partner at the other side) is \emph{not} defined exogenously, but relies on their consulting for time and power.

\revj{Based on the above discussion, we address the following matching problem for the proposed market:}
\begin{question}[Relay-Assignment]\label{ques:1}
What kind of matching pairs will emerge from users' \revj{strategic} interactions?
\end{question}
\begin{question}[Resource-Exchange]\label{ques:2}
What is the detailed resource exchange (i.e., the access time and relay power) for each pair of matched PU and SU?
\end{question}

\subsection{Market Equilibrium  -- Stable Matching}\label{sec:model:eq}

%Now we consider what kind of matchings will actually emerge
Now we consider the above matching problem from a game-theoretical analysis.
\revv{Note that game theory is applicable to our model, as PUs and SUs are rational and self-interested, and always want to maximize their own utilities. Furthermore,  one user's decision will affect others' utilities and decisions.}
To solve the matching problem, it would be useful to understand what kind of matching will be ``broken'' (and thus unstable).

\begin{table*}[t]
\vspace{-2mm}
\small
\centering
    \caption{Information Scenarios and Associated Market Equilibria. }\label{tbl:eq}
    \vspace{2mm}
 \begin{tabular}{|c|c|c|c|c|}
\hline
 \textbf{Information Scenario} &
 \textbf{Each PU $m$'s Knowledge} &
\textbf{Equilibrium} &
\textbf{Property} &
\textbf{Section No.} \\
\hline
\hline
Complete
&
$\{G_m, G_{n(m)}, G_{m,n}, G_{n,m}\}_{ n\in\N, m\in \M}$
&
{\POEQ}
&
Pareto-optimal for PUs
&
Section \ref{sec:main}
\\
\hline
\hline
Incomplete
&
$\{G_m, G_{n(m)}, G_{m,n}, G_{n,m}\}_{ n\in\N}$
&
{\RBEQ}
&
Guaranteed for PUs
&
Section \ref{sec:main2}
\\
\hline
%\multirow{3}*{SU-proposal} &  Complete & SU-dominant $ {Eq.}$ & Optimal for all SUs \\
%\cline{2-4}
% & Weakly Incomplete &  SU-Semidominant $ {Eq.}$ & Bounded by SU-pessimistic $ {Eq.}$ \\
%\cline{2-4}
% & Strongly Incomplete  & SU-recessive $ {Eq.}$ & Bounded by SU-pessimistic-by-guess $ {Eq.}$ \\
%\hline
\end{tabular}
\vspace{-4mm}
\end{table*}

\begin{figure}[tt]
\centering
\includegraphics[width=0.3\textwidth]{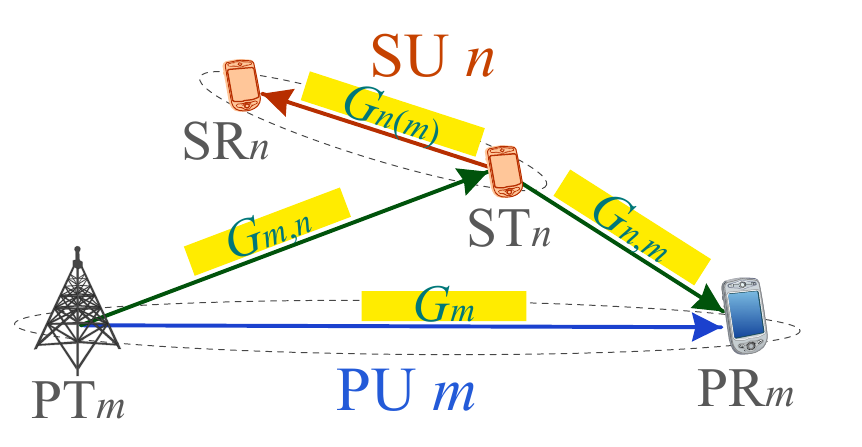}
\caption{Illustration of the key network information.}
\label{fig:network-channel}
\vspace{-2mm}
 \end{figure}

\begin{definition}[Unstable Matching] \label{def:broken-matching}
A matching will be \emph{broken} if one of the following conditions is satisfied:
\begin{enumerate}
\item[(a)] There exist a PU and an SU who are not matched to each other but both prefer to be;
\item[(b)]  There exists a PU or an SU with a negative utility in the existing matching.
\end{enumerate}
\end{definition}
In (a), the PU and the SU would discard their current partners and match to each other. In (b), the PU or SU would remain single rather than stick to the current partner. \revj{Thus, a broken matching is ``unstable''.}

\rev{As long as a matching is not broken, we call it a \emph{stable matching}, or equivalently, a market \emph{equilibrium}.}

\begin{definition}[Market Equilibrium]\label{def:eq}
A market equilibrium is a stable matching, where no user can improve its utility via unilateral  deviation (i.e., choosing another partner or changing resource exchange details).
%\footnotesc{\revh{Note that the stable matching in a two-sided market is a bit different from the Nash equilibrium (NE) in a game. NE just requires no profitable ``unilateral'' deviation, while stable matching requires no profitable deviation, even by multiple users at the same time.}}
%\com{This is a bit different from the NE concept. NE just requires no profitable ``unilateral'' deviation. However, stable matching requires no profitable deviation, even by multiple users at the same time? Need to explain this point.}\com{Lingjie: Though one individual's deviation needs its present or future partner's agreement, it seems that stable matching is still an equilibrium. We may not to mention this.}
\end{definition}

%Based on Definition \ref{def:eq},
Based on the above definitions, we can rewrite the matching problem (given in Questions \ref{ques:1} and \ref{ques:2})  into the following equivalent question:
%can be transformed into the following equivalent question:
%\begin{question}\label{ques:3}
 \emph{what market equilibrium (stable matching) will emerge in the two-sided matching market?}
% \end{question}
\revv{Notice that there are many important factors affecting the   market equilibrium realization, one of which is:  which side of the market has the market power to propose cooperation offers to the other side \cite{gale1962college}.
%As shown in \cite{gale1962college}, there \revj{may exist} multiple equilibria in a two-sided matching market, and what kind of equilibria will emerge depends largely on which side of the market has the power to initially propose offers.
In this paper, we consider a \textbf{{PU-proposal market}}, where PUs have the authority to propose offers.\footnote{{Note that the analysis for the PU-proposal market \textbf{can be directly extended to the SU-proposal market}, where SUs propose offers to PUs, and PUs simply accept or reject the received offers.}}
That is, each PU decides which SU to select and what time reward and relay power request to/from \rev{that} SU, and each SU passively accepts or rejects the received offer.
The study of a PU-proposal market is \rev{motivated} by the fact that PUs usually have more market power than SUs in \rev{cooperative spectrum sharing}, since their number is limited and they own the scarce spectrum licenses.}

\subsection{Network Information}\label{sec:model:info}

\begin{figure}[tt]
\centering
 \includegraphics[width=0.45\textwidth]{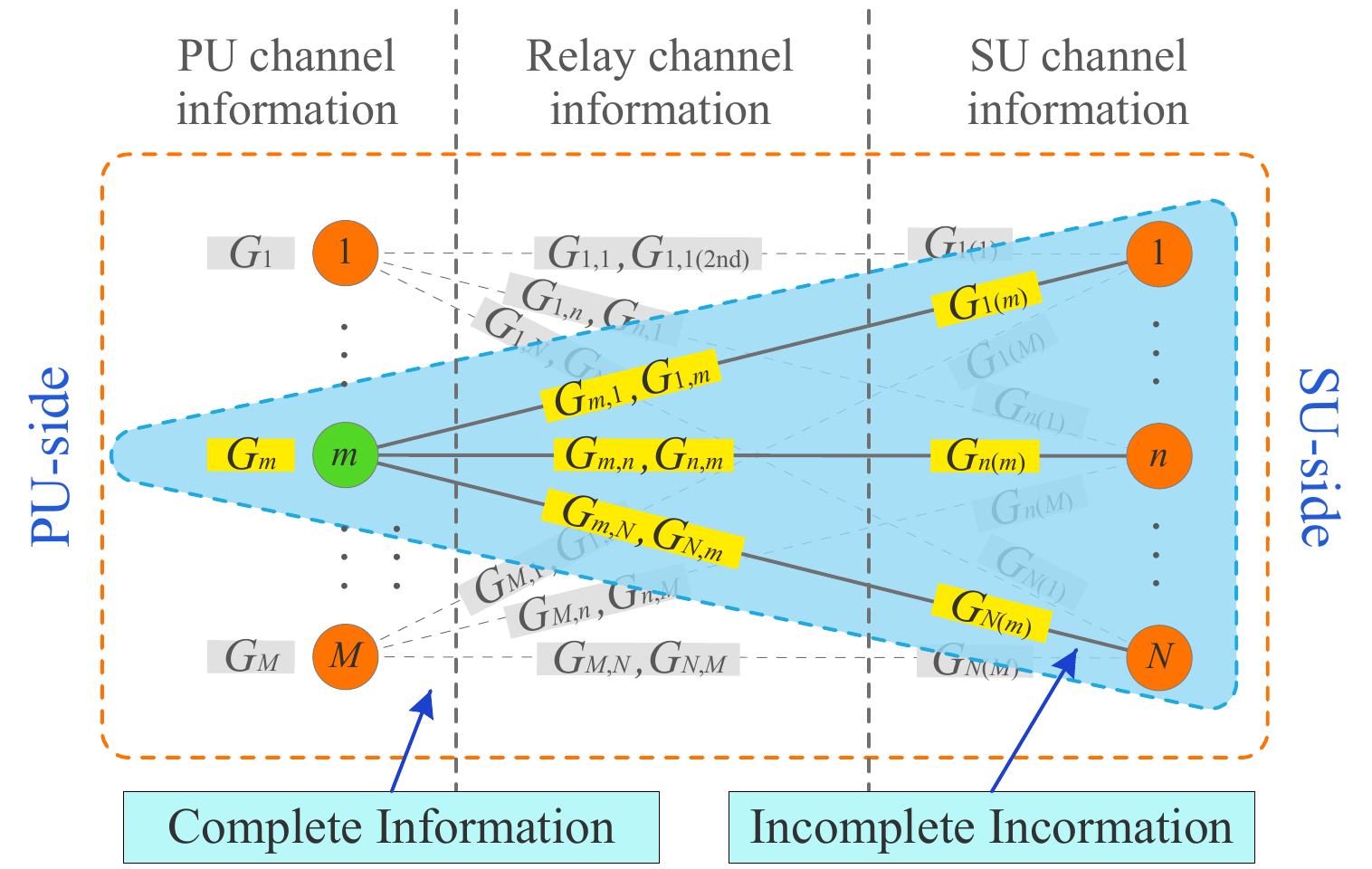}
\caption{Illustration of complete and incomplete information scenarios.
Under complete information, each PU $m$ knows all   network information (in the rectangle).
Under incomplete information, each PU $m$ knows its local information (in the blue triangle). }
\label{fig:information}
\vspace{-3mm}
\end{figure}

%\com{do not use non-standard abbreviations in the main text. I have changed it.}

In addition to who are the proposers, the information scenario also affects the equilibrium.
That is, which equilibrium will eventually appear also depends on how much network information are known to PUs.
%{We first list the key network information as follows.}
In our model, the key network information mainly include\footnote{Notice that we view the cooperative transmission time $T_m$ of each PU $m$ and the power sensitivity $C_n$ of each SU $n$ as public information. {This is because these parameters are usually pre-defined and keep unchanged in a long time, given the types of PUs or SUs.}}
%It is notable that our analysis is directly applicable as we consider them, as private information.}
%
%We consider three  information scenarios: (i) \textbf{{Complete}}  \rev{information}, where each PU knows all network information, (ii) \textbf{{Weakly incomplete}} \rev{information}, where each PU knows all the information related to himself, but not those between others, and (iii) \textbf{{Strongly incomplete}} \rev{information}, where each PU knows only its own information, but not  the information related to himself.
%\com{Lingjie: the last two short sentences seem opposite to each other and confusing. Also, it seems we can delete this paragraph seems we mention the same thing every soon.}
%
%\revh{More precisely, we list the key information as follows.}
\begin{itemize}

\item[(a)] $G_m$: the direct channel gain of each PU $m$;

\item[(b)] $G_{n(m)}$: the direct channel gain of each SU $n$ on each PU $m$'s spectrum band;

\item[(c)] $G_{m,n}$ and $G_{n,m}$: the relay channel gains  between each pair of PU $m$ and SU $n$.
\end{itemize}
 %Without any loss of generality and for the sake of simplicity, we assume $T_m$ and  $C_n$ are fixed and therefore public for all users. Thus, we focus on the information of channel gains: $G_m$, $G_n(m)$, $G_{m,n}$ and $G_{n,m}$, $\forall m,n$.
For convenience, we illustrate the network information (between PU $m$ and SU $n$) in Figure \ref{fig:network-channel}, where the label in each link denotes the associated channel gain.

\revv{In this work, we consider two different information scenarios: \emph{complete} information and \emph{incomplete} information, depending on how much network information the PUs know.
%, and \emph{strongly incomplete information}.
Specifically, in the complete information scenario, each PU $m$ is assumed to know the whole network information, i.e., $\{G_m, G_{n(m)}, G_{m,n}, G_{n,m}\}_{ n\in\N, m\in \M}$.
In the incomplete information scenario, each PU $m$ is assumed to know its local network information only, i.e., $\{G_m, G_{n(m)}, G_{m,n}, G_{n,m}\}_{n\in \N}$, but not those of other PUs.
Note that the incomplete information scenario is more practical, as a PU can use reference signal strength (RSS) to detect the channel gains to all potential SUs, yet it is hard for him to estimate other PUs' channel gains.
For convenience, we summarize these information scenarios in Figure \ref{fig:information}.\footnote{\revmm{We further study a \textit{strongly incomplete} information scenario  in the online technical report \cite{techrpt}, where each PU $m$ does not even know each SU $n$'s channel gain, i.e., $G_{n(m)}$. The equilibrium analysis for such a stronger incomplete information scenario is similar.}}
%\revmm{We will further discuss how PUs obtain these informations in Appendix-B.}

In what follows, we will characterize the sufficient and necessary conditions for the market equilibrium (of the PU-proposal market) in Section \ref{sec:main}, and study which equilibria will emerge in complete and incomplete information scenarios in Section \ref{sec:main2}.
%Specifically, in the complete information scenario, we show that the unique Pareto-optimal equilibrium will always emerge, and we further characterize this Pareto-optimal equilibrium (denoted by {{\POEQ}}).
%In the incomplete information scenario, it is difficult to characterize the detailed equilibrium due to the difficulty in characterizing the SUs' mis-representation behaviors.
%Thus, we characterize the robust equilibrium (denoted by {{\RBEQ}}), which provides every PU a guaranteed utility under all possible mis-representation behaviors of SUs.
For convenience, we summarize these information scenarios and the associated market equilibria in Table \ref{tbl:eq}.}

\section{A Simplified Model}\label{sec:toy}

Before studying the market equilibrium in the general model, we first consider a simplified model where the access time reward offered by each PU (in Phase III) and the relay power contributed by each SU (in Phase II) are fixed. Thus, we only need to consider the binary matching between PUs and SUs (i.e., the relay assignment problem defined in Question \ref{ques:1}).\footnotesc{This model is similar to the two-sided market model in \cite{gale1962college,roth1984misrepresentation}, and our results in this section also refer to those in \cite{gale1962college,roth1984misrepresentation}.}

Note that we use this simple model to illustrate some key properties of the proposed matching market, \revh{which is essential for understanding the general model in Sections~\ref{sec:main} and \ref{sec:main2}. \revj{Moreover, this simplified model is useful for the network scenario with inflexible relay protocols or static physical layer specifications, where secondary access time and relay power cannot be freely adjusted.}}   %{In Section~\ref{sec:main}, we will further discuss the general matching market with variable time and power. }

%\response{by Lingjie}
%\revh{In this section, we study a simple model where a PU $m$ fixes its time allocation to any relaying SU to be the same (denoted by $\hat{t}_m$) and an SU $n$ fixes its power to relay any PU's traffic to be the same (denoted by $\hat{p}_n$). We will use this model to illustrate some key properties of the matching market. These properties help readers understand the general model in Section~\ref{sec:main}, where time and power are users' decision variables and may be different from pair to pair. }

\subsection{\revj{Users' Preferences}}

%When time and power are fixed, we only need to consider the relay assignment problem in Question \ref{ques:1}.
%\revj{Under \textbf{complete information}}, a PU knows its exact utility for choosing each of the SUs, \revj{and thus the} PU's preferences on all SUs are fixed. Same applies to the preferences of an SU for all the PUs.
%The problem thus}
\revxx{With fixed time reward and relay power, each user's utility for each particular partner on the other side is fixed.}
Thus,  {the model} degenerates to a two-sided matching problem \emph{without} utility transferring (similar to the Marriage model considered in \cite{gale1962college}).
%Thus, we only need to consider the relay assignment problem in Question \ref{ques:1}, i.e., what kind of matching will result from.
%
%\response{by Lingjie}
%\rev{When time and power are fixed, we only need to consider the relay assignment problem in Question \ref{ques:1}. \com{In this simple model,a PU knows its exact utility for choosing each of the SUs. that a PU's preferences on all SUs are fixed. Same applies to the preferences of an SU for all the PUs.} \com{Lingjie: This means: This is not true with incomplete info. PU in toy and general models know its preferences for SUs as long as with complete info.}\revh{If a PU and an SU decide to cooperate, their utilities are fixed.} The two-sided market thus}
%%or SU's utility on a particular partner (at the other side), given by (\ref{eq:PU_ut}) or (\ref{eq:SU_ut}), is also fixed. Accordingly, each user's preferences on all possible partners are defined \emph{exogenously}, and the problem
%degenerates to a \com{two-sided matching }problem \emph{without} utility transferring (similar to the Marriage model in \cite{gale1962college}).
%%Thus, we only need to consider the relay assignment problem in Question \ref{ques:1}, i.e., what kind of matching will result from.

Let ${t}_m$ denote the (fixed) spectrum access time rewarded by PU $m$ for any SU collaborating with him, and ${p}_n$ the (fixed) relaying power offered by SU $n$ for any PU collaborating with him.
%\rev{We let ${t}_m$ denote the (fixed) time that PU $m$ assigns to a cooperative relay SU}, and let ${p}_n$ denote the  (fixed) power that  SU $n$ offers\rev{to relay any PU's traffic. }
%%For every pair of PU $m$ and SU $n$, we can derive the PU $m$'s utility by Eq. (\ref{eq:PU_ut}) and the SU $n$'s utility by Eq. (\ref{eq:SU_ut}).
The preferences of a PU $m$ can be represented by an ordered list in the \rev{decreasing} preference, $P(m)$, on the SUs,\footnotesc{\rev{The number of elements in set $P(m)$} could be smaller than $N$, that is, PU $m$ \rev{is only interested in a subset} of SUs.}
\begin{equation}\label{eq:prefer-pu}\textstyle
P(m) = \{n_1,n_2,...,n_p\},
\end{equation}
subject to (i) $\Pi_{n_i}^m({t}_m,{p}_{n_i}) \geq \Pi_{n_j}^m({t}_m,{p}_{n_j})$ if $i<j$,
%\footnotesc{We assume that \rev{there is no tie}, which is reasonable for the network with continue information.}
and (ii) $\Pi_{n}^m({t}_m,{p}_{n}) \geq 0$ \revj{for any SU} $n \in P(m)$, and $\Pi_{n'}^m({t}_m,{p}_{n'}) < 0$ for any SU $n'\notin P(m)$. Here $\Pi_n^m(\cdot)$ is the PU $m$'s utility defined in (\ref{eq:PU_ut}).
The first condition means that PU $m$ prefers an SU \rev{who provides him a} higher utility, and the second condition means that PU $m$ never selects an SU \rev{who brings him a} negative utility.~~~~~~
%\com{How the PU knows this preference list under incomplete info? May need to explain here.}\response{I suggest that we shall not discuss the information issue here. Actually, the preference is an objective parameter, which is independent to the users' information.}

Similarly, the preference list of an SU $n$ on the PUs can be represented by
\begin{equation}\label{eq:prefer-pu}\textstyle
Q(n) = \{m_1,m_2,...,m_q\},
\end{equation}
subject to (i) $\Delta_{n}^{m_i}({t}_{m_i},{p}_{n}) \geq \Delta_{n}^{m_j}({t}_{m_j},{p}_{n})$ if $i<j$, and (ii) $\Delta_{n}^{m}({t}_{m},{p}_{n}) \geq 0 $ if $m \in Q(n)$, and $\Delta_{n}^{m'}({t}_{m'},{p}_{n}) < 0 $ $m' \notin Q(n)$. Here $\Delta_n^m(\cdot)$ is the SU $n$'s utility in (\ref{eq:SU_ut}).

\subsection{\revj{An Example and Its Equilibrium}}

We provide an example to illustrate the equilibria \revj{under the fixed preferences}.\footnote{The complete equilibrium analysis for such a simplified model (with the fixed preferences) can be referred to \cite{techrpt}.}
\revj{Recall that}  a binary matching between PUs and SUs is represented  by a correspondence $\mu(\cdot)$, where $\mu_m=n$ (or equivalently, $\mu_n = m$) means PU $m$ is matched to SU $n$, and $\mu_m = \emptyset$ or $\mu_n = \emptyset$ means PU $m$ or SU $n$ remains single (i.e., not engaged in the cooperative spectrum sharing).% \revh{characterized by Lemma~\ref{lemma:necesuff-toy}.}
\begin{example}\label{examp:1}
Consider a two-sided market with 3 PUs ($m_1,m_2,m_3$) and 3 SUs ($n_1,n_2,n_3$) with preferences:

\vspace{1mm}
\noindent
\begin{tabular}{ c|c||c|c }
\hline
\multicolumn{2}{ c||} {~~~~~~~\emph{PUs' preferences}~~~~~~}& \multicolumn{2}{c } {~~~~~~~\emph{SUs' preferences}~~~~~~} \\
\hline
$P(m_1)$ & $n_1,\ n_2,\ n_3$ & $Q(n_1)$ & $m_2,\ m_3,\ m_1$ \\
\hline
$P(m_2)$ & $n_2,\ n_3,\ n_1$ & $Q(n_2)$ & $m_3,\ m_1,\ m_2$ \\
\hline
$P(m_3)$ & $n_3,\ n_1,\ n_2$ & $Q(n_3)$ & $m_1,\ m_2,\ m_3$ \\
\hline
\end{tabular}
\vspace{1mm}

According to the definition of equilibrium in Definition~\ref{def:eq}, we can easily obtain the following three equilibria $\mu^a$, $\mu^b$, and $\mu^c$ for the above example:

\vspace{1mm}
\noindent
\begin{tabular}{ c|m{3mm}|m{3mm}|m{3mm}||m{3mm}|m{3mm}|m{3mm}||m{3mm}|m{3mm}|m{3mm} }
\hline
\emph{Equilibria} & \multicolumn{3}{c||} {$\mu^a(\cdot)$}  & \multicolumn{3}{c||}  {$\mu^b(\cdot)$}  & \multicolumn{3}{c } {$\mu^c(\cdot)$}  \\
\hline
\emph{PUs} & $m_1 $ & $ m_2 $&$ m_3  $&$ m_1 $&$ m_2  $&$ m_3  $&$ m_1  $&$ m_2 $&$  m_3 $ \\
\hline
\emph{SUs} & $n_1 $ & $ n_2 $ & $ n_3 $&$ n_2  $&$ n_3  $&$ n_1 $&$ n_3  $& $ n_1  $ & $ n_2$\\
\hline
\end{tabular}
\vspace{1mm}

%Obviously, there does not exist any individual or pair of individuals, who would deviate from the current matching if one of the above equilibria is achieved.
%Obviously, in all of the above equilibria, there does not exist a PU and an SU who are not matched to each other but both prefer to be.
\revv{Let us check the feasibility of the above equilibria using the definition of market equilibrium.
Take the equilibrium $\mu^b$ as an example, where PU $m_1$ is matched to SU $n_2$, PU $m_2$ is matched to SU $n_3$, and PU $m_3$ is matched to SU $n_1$.
%, and
%consider a pair of unmatched PU and SU, say, SU $n_1$ and PU $m_1$.
%For SU $n_1$, we have: $\Delta_{n_2}^{m_1} > \Delta_{n_2}^{m_2}$, which implies that SU $n_2$ prefers PU $m_1$ than its partner PU $m_2$.
We can easily find that PU $m_1$ prefers SU $n_1$ than its current partner SU $n_2$ (as $\Pi_{n_2}^{m_1} < \Pi_{n_1}^{m_1}$), but SU $n_1$ has no incentive to discard its current partner PU $m_3$ and pair with PU $m_1$ (as $ \Delta_{n_1}^{m_3} > \Delta_{n_1}^{m_1}$).
Similarly, PU $m_2$ prefers SU $n_2$ than its current partner SU $n_3$, but SU $n_2$ has no incentive to discard its current partner PU $m_1$ and pair with PU $m_2$. PU $m_3$ prefers SU $n_3$ than its current partner SU $n_1$, but SU $n_1$ has no incentive to discard its current partner PU $m_2$ and pair with PU $m_3$.
Thus, there does not exist a pair of PU and SU who are not matched to each other but both prefer to be.}
%Thus, PU $m_1$ will not break the current matching, which further implies that SU $n_2$ will not break the current matching as it knows that PU $m_1$ will not accept its matching request).
%This implies that even if $n_2$ prefers $m_1$ than its partner $m_2$, he will not deviate to $m_1$, since $m_1$ will not accept him.
\hfill
$\square$
\end{example}

%A following essential question is: \emph{Do equilibria always exist?} The answer is YES. Specifically,
%\begin{lemma} [Existence]
%An equilibrium exists for any two-sided PUs-SUs market with fixed time and power.
%\end{lemma}
%The detailed proof can be referred to \cite{gale1962college}, where Gale and Shapley demonstrated a ``\emph{deferred acceptance}'' algorithm which effectively converges to an equilibrium.
%
%The basic idea of the deferred acceptance algorithm is as follows: At any round, each PU proposes to its most preferred SU among those who have \emph{not yet} rejected him, and each SU who receives multiple proposals rejects all but its most preferred.
%%To start, each proposer (e.g., PUs in the PU-proposal market) proposes to its favorite partner, and each responser (e.g., SUs in the PU-proposal market) who receives multiple proposals rejects all but her most preferred. At any following step, each proposer who was rejected at the previous step proposes to its next choice (i.e., to its most preferred partner among those who have not yet rejected him), and each responser rejects all but its most preferred among the group consisting of the new proposers together with any partner it may have kept engaged from the previous step.
%It is worth noting that with different users as proposers, the resulting equilibria may be different with each other. More concisely, the equilibrium resulting from PUs proposing may be different to that from SUs proposing. We show this by the following example.

%!TEX root = CoopSpecShar_main.tex
%SourceDoc CoopSpecShar_main.tex

\section{Market Equilibrium Analysis}\label{sec:main}

\revj{Now we characterize the market equilibrium in a general model where the access time reward and the relay power} \revj{are {not} fixed}.
In this case, each user's utility \rev{(with a particular partner at the other side) is no longer fixed, but changes with the resource exchange (i.e., the access time reward and relay power)  between them.
Accordingly, the preference list of each user depends on its current resource exchanges with users at the other side.}
Thus, we need to consider not only the binary matching between PUs and SUs (i.e., the relay assignment problem defined in Question \ref{ques:1}), but also the resource exchange between each pair of matched users (i.e., the resource exchange problem defined Question \ref{ques:2}).
%\revv{This is one of the key differences between this work and the previous literature on two-sided matching (e.g., \cite{gale1962college,roth1984misrepresentation}).}
% \com{(i.e., determining the time and power for each pair of matched PU and SU)} defined \revh{in Section \ref{sec:model}.D.}} \com{in Section ?}

In what follows, we will start by defining the \emph{utility transfer function}, which captures the interaction \revh{within each matching pair}. Then we characterize the necessary and sufficient conditions for market equilibrium.
We will show that there are multiple (infinite number) of market equilibria.
In the next section, we will further study which specific equilibrium will actually emerge in different information scenarios.
%We further characterize the unique optimal equilibrium for PUs under complete information, and the pessimistic equilibrium for PUs under weakly incomplete information \rev{(as an achievable lower-bound of PUs' utilities)}. Finally, we study the equilibrium under strongly incomplete  information.

\subsection{Utility Transfer Function - UTF}

\rev{Suppose that PU $m$ is matched to SU $n$.}
We first study how their utilities change with the access time reward and the relay power.
%The first question is how they decide the time and power.
%\com{I remove the following sentence ``in the presence of other users'', since we'd better not involve other users into this section. Just view UTF as a function which derive the optimal PU's utility given the SU's required utility.}.

Recall in the PU-proposal market, PUs propose the offers and SUs decide whether to accept or not.
Let $\delta_n^m$ denote the SU $n$'s highest utility from all other PUs' offers (except PU $m$'s) and from remaining single  (not matched to any PU), called \emph{reservation utility} of SU $n$ (towards PU $m$).\footnote{\revv{Note that an SU $n$'s reservation utility is PU-dependent, i.e., $\delta_n^{m_1}$ may be different from $\delta_n^{m_2}$, as it is the SU's highest utility from all but one of the PUs.}}
Then, to attract SU $n$'s cooperation, PU $m$ has \rev{to offer SU $n$ a utility no less than $\delta_n^m$}.
%Let $\pi_m$ and $\delta_n$ denote the utilities of PU $m$ and SU $n$, respectively. %, which is defined as the maximal utility an user can achieve from outside option.
%Obviously, to keep the matching, both PU $m$ and SU $n$ must achieve a utility, under certain time and power (termed as an offer), no less than $\pi_m$ and $\delta_n$, respectively.
%In the PU-proposal market,
%\rev{Thus, to maximize its utility, PU $m$ needs to solve the following optimization problem:}
Thus, the PU $m$'s maximum utility is given by
%\footnotesc{\rev{The optimization (\ref{eq:opt_pu}) implies that PU $m$ knows enough network information (e.g., in the complete information scenario).}}
\begin{equation}\label{eq:opt_pu}
\begin{aligned}
\Pi_n^{m*}   =  & \max_{\{p_n^{m }, t_n^{m }\}}  \quad \Pi_n^{m}(p_n^{m }, t_n^{m }) \\
& \textstyle s.t.  \quad \Delta_{n}^m(p_n^{m }, t_n^{m }) \geq \delta_n^m,
\end{aligned}
\end{equation}
where $\Pi_n^{m}(p_n^{m }, t_n^{m })$ is PU $m$'s utility defined in (\ref{eq:PU_ut}),
$\Delta_{n}^m(p_n^{m }, t_n^{m })$ is  SU $n$'s utility defined in (\ref{eq:SU_ut}).
\revv{Notice that the constraint must be tight at the optimality. This implies that the variable $t_n^{m }$ (or $p_n^{m }$) can be rewritten as a function of $p_n^{m }$ (or $t_n^{m }$), and thus the problem (\ref{eq:opt_pu}) can be transformed into a single-variable  optimization problem. Hence,   problem (\ref{eq:opt_pu}) can be solved efficiently using many one-dimension  exhaustive search methods. We denote the computational complexity of solving (\ref{eq:opt_pu}) as $\rho $.}~~~~~~~

It is easy to check that the \rev{PU $m$'s maximal utility} $\Pi_n^{m*} $ given by (\ref{eq:opt_pu}) is strictly decreasing in \revj{SU $n$'s reservation utility} $\delta_n^m$. %Since $\Pi_n^{m*}$ is a function of $\delta_n$, we can rewrite  it as
\rev{Thus, we can write} $ \Pi_n^{m*}$ as a decreasing function of $\delta_n^m$:
$$\Pi_n^{m*} \triangleq f_n^{m} (\delta_n^m).$$
%The increase of SU $n$'s utility decreases the PU $m$'s maximal utility, and vice versa. In this sense, we
We refer to $  f_n^{m} (\cdot)$ as the {\textbf{Utility Transfer Function}} (\textbf{UTF}).
%Formally,
%\begin{definition}[UTF]
%The utility transferring function $  f_n^{m} (\delta_n)$  is given by optimization problem (\ref{eq:opt_pu}).
%\end{definition}
Furthermore, we denote
$$
g_n^{m}(\cdot) \triangleq  {f_n^{m}}^{(-1)}(\cdot)
$$ as the \textbf{Inverse Utility Transfer Function} \textbf{(IUTF)}, i.e., $ \pi_n^m = f_n^{m}(\delta_n^m) $ if and only if $ \delta_n^m = g_n^{m}(\pi_n^m)$.
Note that $f_n^{m}(\delta_n^m) $ denotes the PU $m$'s \emph{maximal} utility when giving SU $n$ a utility $\delta_n^m$, and  $g_n^{m}(\pi_n^m)$ denotes the SU $n$'s \emph{maximal} utility when giving PU $m$ a utility $\pi_n^m$.\footnote{An illustration of UTF is provided by Figure \ref{f-simu1} in \cite{techrpt}.}
%Figure \ref{f-simu1} in the simulations illustrates the UTF.

\revmm{
To facilitate the understanding of the later analysis, we list some important notations in Table \ref{tbl:notation2}.}

\begin{table}[t]
\small
    \caption{Important Notations in Section \ref{sec:main}}\label{tbl:notation2}
    \hspace{-2mm}
\begin{tabular}{m{0.029\textwidth}m{0.43\textwidth}}
\hline
\hline
 $f_n^{m}(\cdot)$ &   PU $m$'s  {maximal} utility given the  matched SU's utility;\\
 $g_n^{m}(\cdot)$ &   SU $n$'s  {maximal} utility given the  matched PU's utility;\\
 \hline
 $\mu_n$ &   The PU matched to SU $n$;\\
 $\mu_m$ &   The SU matched to PU $m$;\\
 $\delta_n$ & The SU $n$'s achieved utility in a given  matching;\\
 $\bar{\delta}_n$ &  The SU $n$'s highest achievable utility in a matching;\\
 $\underline{\delta}_n$ &  The SU $n$'s lowest acceptable utility in a matching;\\
\hline
\hline
\end{tabular}
\vspace{-4mm}
\end{table}

%For notational convenience, we further denote
%$(\pi_n^{m})^o =  f_n^{m}(0)  $ as the maximum \rev{achievable utility of} PU $m$ when leaving zero utility to SU $n$, and $(\delta_n^m)^o  = {f_n^{m}}^{(-1)}(0) = g_n^{m}(0)$ as the maximum \rev{achievable utility of} SU $n$ when leaving zero utility to PU $m$.
%Without loss of generality, we assume $(\pi_n^{m})^o \geq 0$ and $(\delta_n^m)^o\geq 0, \forall m,n$, i.e., the cooperation between any PU and SU  brings certain benefit.

\subsection{Sufficient \& Necessary Conditions}

Now we study the  sufficient and necessary conditions for an
equilibrium.
%Without loss of generality, we assume $M=N$, i.e., the same number of PUs and SUs. Note that for a general case with $M\neq N$, we can simply introduce certain dummy users to let $M=N$. Obviously, for a dummy SU $n$, we have $f_n^{m} (\delta_n) \equiv 0, \forall \delta_n$, since a PU $m$ matched to a dummy SU actually means PU $m$ remains single. Similarly, for a dummy PU $m$, we have ${g_n^{m}} (\pi_m) \equiv 0, \forall \pi_m$.
\rev{The optimization problem (\ref{eq:opt_pu})} shows that the PU $m$'s maximal utility depends on \rev{SU $n$'s  utility $\delta_n^m$}, so does \rev{the optimal access time $t_n^m$ and relay power $p_n^m$}.
%the PU $m$'s maximal utility $\pi_n^{m*}$ and the best offer (time and power) are totally determined by the utility for SU $n$, which is equivalent to (or strictly, tiny bit higher than) the SU $n$'s reserve utility $\delta_n$.
Thus, the resource exchange problem in Question \ref{ques:2} is equivalent to the following utility division problem: \emph{what is the utility for each SU in a matching?} \rev{This means that we can rewrite a matching as}% Thus, a  matching is generalized to
\begin{equation}
 \{(\mu_n, \delta_n),\ \forall n \in \mathcal{N} \},
\end{equation}
where $\mu_n$ denotes the PU matched to SU $n$, and $\delta_n$ denotes the SU $n$'s utility in the given matching.\footnotesc{Note we omit the superscript $\mu_n$ in $\delta_n^{\mu_n}$ for simplicity of writing.}
%In the following, \textbf{we will write $\mu(n) $ as $\mu_n$ for simplicity.}
Obviously, in such a matching,  the utility of PU $ \mu_n$ (i.e., the PU matched to SU $n$) is $f_n^{\mu_n}(\delta_n)$, \revv{which is   PU $\mu_n$'s maximum achievable utility computed by (\ref{eq:opt_pu}).}
Note that the utility is zero for an unmatched SU $n$ (with $\mu_n=\emptyset$) or PU $m$ (with $m \neq \mu_n, \forall n \in \mathcal{N}$).

\revv{In what follows, we will present the lower-bound and upper-bound of each $\delta_n$ (for the matched PU $\mu_n$ and SU $n$), such that none of PUs and SUs has an incentive to break the current matching.
To achieve this, for each pair of matched PU $\mu_n$ and SU $n$: (i) PU $\mu_n$ (or SU $n$) has no incentive to break the current matching by remaining single, or by pairing with another SU (or PU), and (ii) all PUs other than $\mu_n$ (or all SUs other than $n$) have no incentives to discard their respective partners and pair with SU $n$ (or PU $\mu_n$).}

%The discussions in the left column on this page can be better organized. At this moment, there are many notations, and a reader needs to take a lot of efforts following through the notations in order to arrive at the final conclusion in Lemma 5.
%
%I am not sure if there is a way to simplify the notations. But at least we should be able to provide a roadmap to the readers before going through the IR, IC, and CC constraints. In particular, after the paragraph where we define (12), we can tell the readers that the purpose of the discussions is to drive the lower-bound and upper-bound of an SU's achievable utility in a matching.

According to Definition~\ref{def:eq}, we first have the following necessary conditions for an equilibrium.
\begin{lemma}\label{lemma:nece-row}
If $\{(\mu_n, \delta_n), \forall n \}$ is an equilibrium, then the following conditions hold: for each SU $n\in \mathcal{N}$,
\begin{itemize}
\item[] (IR)~~$f_n^{\mu_n}(\delta_n) \geq 0$;
\item[] (IC)~~$f_n^{\mu_n}(\delta_n) \geq  f_k^{\mu_n}(\delta_k),\quad{\forall k \neq n}$,
\end{itemize}
where $\mu_n$ denotes the PU matched to SU $n$.
\end{lemma}

The first condition is generally referred to as \emph{Individual Rationality} (IR), and the second condition is generally referred to as \emph{Incentive Compatibility} (IC).
The proof follows directly from the definition of equilibrium: If the IR condition is violated by a PU $\mu_n$, then PU $\mu_n$ will discard its partner (SU $n$) and remain single; If the IC condition is violated by a PU $\mu_n$, i.e., there exists another SU $k\neq n$ such that $f_n^{\mu_n}(\delta_n) <  f_k^{\mu_n}(\delta_k)$, then PU $\mu_n$ \rev{can get a better utility by paring up with SU $k$ instead.}\footnotesc{\rev{With a proper offer from PU $\mu_n$ \revh{(e.g., giving SU $k$ a utility $\delta_k+\epsilon$, where $\epsilon$ is any infinitesimal positive number)}, SU $k$ will also improve its utility through the new pairing and thus will accept.}}

%Now, Let's analyze the strategy of a particular PU $m$.
%Note the necessary conditions in Lemma \ref{lemma:nece-row} are rough.
%Next we generalize the more concise necessary conditions. \com{not sure what previous sentence means.}

The IC and IR conditions for a PU $\mu_n $ ensure that PU $\mu_n $ has no incentive to change its decision.
We further introduce the \emph{Competitive Compatibility} (CC) condition for PU $\mu_n $, which ensures that no other PUs has the incentive to compete with a PU $\mu_n $ (for SU $n$).
Intuitively, the CC condition for a PU $\mu_n$ (i.e., the PU matched to SU $n$) means no other PU $m$ can achieve a higher utility on SU $n$ than on its current partner SU $\mu_m$.
Formally,
%the CC condition for a PU $\mu_n$ is:\footnotesc{Note that $f_{\mu_m}^m(\cdot) \equiv 0 $ if $\mu_m = \emptyset$, which denotes the utility of PU $m$ when not engaging in any collaboration.}
%\begin{equation*}
%\begin{aligned}
%\mbox{(CC)~~}
%\textstyle  f_{\mu_m}^m(\delta_{\mu_m}) \geq f_n^{m}(\delta_n),\quad \forall m\neq \mu_n,
%\end{aligned}
%\end{equation*}
%which means no other PU $m$ can achieve a higher utility on SU $n$ than on its current partner SU $\mu_m$.  Here $\mu_m$ denotes the SU matched to PU $m$.
%This ensures all other PUs have no incentive to compete for SU $n$, i.e., PU $\mu_n$ is competitive enough to attract SU $n$. %Obviously, the CC condition for PU $\mu_n$  to PU $m\neq \mu_n$ is just the IC condition for PU $m$   to PU $\mu_n$.

\begin{lemma}\label{lemma:nece-cc}
 If $\{(\mu_n, \delta_n), \forall n \}$ is an equilibrium, then the following condition holds: for each PU $\mu_n\in \mathcal{M}$,\footnotesc{Note that $f_{\mu_m}^m(\cdot)  \equiv  0 $ if $\mu_m = \emptyset$, which denotes the utility of PU $m$ when not engaging in any collaboration.}
\begin{itemize}
\item[] (CC)~~$f_{\mu_m}^m(\delta_{\mu_m}) \geq f_n^{m}(\delta_n),\quad \forall m\neq \mu_n$,
\end{itemize}
where $\mu_m$ denotes the SU matched to PU $m$.
\end{lemma}

The IR condition for a PU $\mu_n$ implies
\begin{equation*}
\textstyle f_n^{\mu_n}(\delta_n) \geq 0 \Rightarrow \delta_n\leq
% (\delta_n^{\mu_n})^o,
g_n^{\mu_n}(0),
\end{equation*}
where $g_n^{\mu_n}(0)  = {f_n^{\mu_n}}^{(-1)}(0)$ is the maximum \rev{achievable utility of} SU $n$ when leaving zero utility to PU $\mu_n$.
The IC condition for a PU $\mu_n$ implies
\begin{equation*}
\begin{aligned}
& \textstyle f_n^{\mu_n}(\delta_n) \geq  f_k^{\mu_n}(\delta_k),\quad  \forall k\neq n
\\
\Rightarrow\ & \textstyle  \delta_n\leq  {g_n^{\mu_n}} \big(f_k^{\mu_n}(\delta_k )\big),\quad  \forall k\neq n
\\
\Rightarrow\  & \textstyle \delta_n\leq \min_{k\neq n} \big\{{g_n^{\mu_n}} \big(f_k^{\mu_n}(\delta_k )\big)\big\},
\end{aligned}
\end{equation*}
where $g_n^{m}(x) =  {f_n^{m}}^{(-1)}(x)$ is the SU $n$'s maximum achievable utility when leaving a utility $x$ to PU $m$.
The CC condition for a PU $\mu_n$ further implies
\begin{equation*}
\begin{aligned}
& \textstyle \delta_n \geq
 g_n^{m}\big(f_{\mu_m}^m(\delta_{\mu_m})\big) ,\quad \forall m\neq \mu_n \\
 \Rightarrow \ & \delta_n\geq
\max_{m\neq \mu_n} \big\{g_n^{m}\big(f_{\mu_m}^m(\delta_{\mu_m})\big) \big\}.
\end{aligned}
\end{equation*}

%Besides, from the IC conditions for all other PUs $m \neq \mu_n$ with respect to PU $\mu_n$ or SU $n$ (i.e., the conditions for PUs $m \neq \mu_n$ not selecting SU $n$), referred to as the \emph{Competitive Condition} (CP) for PU $\mu_n$, we have
%\footnotesc{Note that the set of all users' CC conditions are equivalent to the set of all users' IC and IR conditions, in the sense that the IC condition for PU $\mu_n$ with regard to a PU $\mu_k$, i.e., $f_n^{\mu_n}(\delta_n) \geq f_k^{\mu_n}(\delta_k)$, is just the CC condition for PU $\mu_k$ with regard to PU $\mu_n$.}
%\begin{equation*}
%\begin{aligned}
%\textstyle  f_k^{\mu_k}(\delta_k) \geq f_n^{\mu_k}(\delta_n)
%& \textstyle \Rightarrow \delta_n\geq
% g_n^{\mu_k}\big(f_k^{\mu_k}(\delta_k)\big) , \forall k\neq n \\
% &  \Rightarrow \delta_n\geq
%\max_{k\neq n} \big\{g_n^{\mu_k}\big(f_k^{\mu_k}(\delta_k)\big) \big\}.
%\end{aligned}
%\end{equation*}

Based on the above discussions, we can obtain the \emph{lowest acceptable} utility $\underline{\delta}_n$  and the \emph{highest achievable} utility $\overline{\delta}_n$ for an SU $n$ in an equilibrium $ \{(\mu(n), \delta_n), \forall n \}$, i.e.,
\begin{equation}\label{eq:low}\textstyle
\underline{\delta}_n \triangleq
\max_{m\neq \mu_n} \big\{g_n^{m}\big(f_{\mu_m}^m(\delta_{\mu_m})\big), 0 \big\},
\end{equation}
\begin{equation}\label{eq:high}\textstyle
~~\overline{\delta}_n \triangleq \min_{k\neq n}  \big\{{g_n^{\mu_n}} \big(f_k^{\mu_n}(\delta_k ) \big),
%(\delta_n^{\mu_n})^o
g_n^{\mu_n}(0)
\big\}.
\end{equation}
Intuitively, PU $\mu_n$ cannot offer a utility $\delta_n$ lower than $\underline{\delta}_n$ to SU $n$, otherwise \rev{some} other PU $m\neq \mu_n$ will have the incentive to pair up with SU $n$ (by offering SU $n$ a utility slightly higher than $\underline{\delta}_n$). This will make the original matching broken.
On the other hand, PU $\mu_n$ will never offer a utility $\delta_n$ higher than $\overline{\delta}_n$ to SU $n$, otherwise PU $\mu_n$ would be better off by pairing up with an other SU. This will also make the matching broken.

Therefore, we can \revh{rewrite the necessary conditions in Lemma~\ref{lemma:nece-row} more explicitly as follows:}
%can be formally generalized as follows.
%\begin{lemma}\label{lemma:nece}
If a matching $\{(\mu_n, \delta_n), \forall n \}$ is an equilibrium, the condition holds:
\begin{equation*}
\underline{\delta}_n \leq \delta_n \leq \overline{\delta}_n,\quad \forall n\in \mathcal{N},
\end{equation*}
where  $\underline{\delta}_n  $ and  $\overline{\delta}_n $ are defined in (\ref{eq:low}) and (\ref{eq:high}), respectively.
%\response{by Lingjie}
%\begin{equation}\label{eq:neccgeneral11}
%\underline{\delta}_n \triangleq
%\max_{m\neq \mu_n} \big\{g_n^{m}\big(f_{\mu_m}^m(\delta_{\mu_m})\big), 0 \big\}
%\end{equation}
%and  \begin{equation}\label{eq:neccgeneral22}
%\overline{\delta}_n \triangleq \min_{k\neq n}  \big\{{g_n^{\mu_n}} \big(f_k^{\mu_n}(\delta_k ) \big), (\delta_n^{\mu_n})^o \big\}.
%\end{equation}
%\end{lemma}
%The proof follows directly from Lemma \ref{lemma:nece-row}.
The left \rev{inequality} is due to the CC condition, and the right \rev{inequality} is due to the IC and IR conditions.
%Intuitively, the left inequation ensures that none of other PUs deviates to SU $n$ and the right inequation ensures that PU $\mu_n$ does not deviate to other SU.
%From Lemma \ref{lemma:nece}, we  immediately get the following necessary conditions: $\underline{\delta}_n \leq \overline{\delta}_n ,\forall n\in \mathcal{N}$.

The next theorem shows the above conditions
are not only necessary, but also sufficient.
\begin{theorem}
[\textbf{Sufficient  $\&$ Necessary Condition}]\label{lemma:necsuff1}
A matching $\{(\mu_n, \delta_n), \forall n \}$ is an equilibrium, if and only if
$$
\underline{\delta}_n \leq \delta_n \leq \overline{\delta}_n,\quad  \forall n\in \mathcal{N}.
$$
%where $\underline{\delta}_n  $ and  $\overline{\delta}_n $ are given by  (\ref{eq:low}) and (\ref{eq:high}), respectively.
\end{theorem}

%\begin{proof}
%Please refer to Appendix-D for details.
%\end{proof}
Due to space limit, we put the detailed proof in \cite{techrpt}.

\subsection{Property of Equilibrium}

To facilitate the characterization of equilibrium in different information scenarios (in the next section), we first show the following property for equilibrium.

\begin{lemma}[\textbf{Generalized Lattice Theorem}]\label{lemma:lattice}
Given any possible binary matching $\mu(\cdot)$, if $\{(\mu_n, \delta_n^I), \forall n \}$ and $\{(\mu_n, $ $ \delta_n^{II}), \forall n \}$ are equilibria, then the matching $\{(\mu_n, \delta_n^{X}), \forall n \}$ with $\delta_n^{X} = \min(\delta_n^I,  \delta_n^{II})$ is also an equilibrium.
\end{lemma}

%\begin{proof}
%Please refer to Appendix-E for details.
%\end{proof}

Lemma \ref{lemma:lattice} can be seen as a generalization of the Lattice Theorem proposed by Kunth \cite{Knuth1976Marriage} for the basic two-sided matching (without utility transferring). By repeatedly applying \revj{Lemma \ref{lemma:lattice}} on any two
equilibria, we will reach the \emph{unique} optimal equilibrium for PUs.

\begin{theorem}[\textbf{Optimality}]\label{lemma:opt}
There exists a Pareto-optimal equilibrium $\{(\mu_n, \delta_n^*), \forall n \}$  for PUs, where {every} PU \revj{achieves its maximum utility among} all equilibria.
In addition, $\{(\mu_n, \delta_n^*), \forall n \}$ satisfies:\footnotesc{Similarly, there always exists a Pareto-optimal equilibrium $\{(\mu_n, \delta_n^*), \forall n \}$ for SUs, where \textbf{every} SU achieves its maximum utility among all equilibria. In addition, $\{(\mu_n, \delta_n^*), \forall n \}$ satisfies: $\delta_n^* = \overline{\delta}_n^* , \forall n$.}
$$
\delta_n^* = \underline{\delta}_n^* ,\quad \forall n \in \N.
$$
\end{theorem}

Theorem \ref{lemma:opt} not only shows the existence of the Pareto-optimal equilibrium for PUs, but also suggests a method to calculate the Pareto-optimal equilibrium for PUs. Specifically, given any feasible matching $\mu(\cdot)$, \rev{we can derive} each SU's utility (and the associated time reward and relay power) under the Pareto-optimal equilibrium by jointly solving the following function set:
\begin{equation}\label{eq:funcset}
 \delta_n  = \max_{m\neq \mu_n} \big\{g_n^{m}\big(f_{\mu_m}^m(\delta_{\mu_m})\big), 0 \big\} ,\ \forall n\in \mathcal{N}.
\end{equation}
Obviously, if there is a unique solution for (\ref{eq:funcset}), it must be the unique Pareto-optimal equilibrium. If there is no solution for (\ref{eq:funcset}), the matching $\mu(\cdot)$ is not feasible.

\revv{However, directly solving the above function set is  difficult.
Consider a single equation in the function set
(say, $ \delta_n $).
%$ = \max_{m\neq \mu_n} \big\{g_n^{m}\big(f_{\mu_m}^m(\delta_{\mu_m})\big), 0 \big\}$.
To solve this equation, we first need to compute the utilities of all PUs other than $\mu_n$ on their own matching pairs, i.e.,  $f_{\mu_m}^m(\delta_{\mu_m})$, $\forall m\neq \mu_n$.
Then, we need to compute the SU $n$'s maximum achievable utility when giving each PU $m\neq \mu_n$ a utility $f_{\mu_m}^m(\delta_{\mu_m})$, i.e., $g_n^{m}\big(f_{\mu_m}^m(\delta_{\mu_m})\big)$, $\forall m\neq \mu_n$.
Finally, the SU $n$'s utility $\delta_n$ is the maximum of the above computed maximum achievable utilities and zero.
Obviously, for this single equation, we need to repeatedly solve the optimization problem (\ref{eq:opt_pu})   $2\cdot (M-1)  $ times.
Furthermore, jointly solving $N$ equations makes the problem even  more complicated. This motivates us to study low complexity algorithms to realize different market equilibria in the next section.}

%\comv{How easy to solve the set of equations in (15)? Are they general linear or nonlinear equations? How many equations are there? How is such computation related to the later algorithms proposed?}

\section{Market Equilibrium Realization}\label{sec:main2}

 In the previous section, we characterized the necessary and sufficient conditions of market equilibrium.
In this section, we will study which equilibrium will actually emerge in different information scenarios.

 Specifically, we consider two typical information scenarios: \emph{complete}   and \emph{incomplete} information, depending on how much network information the PUs know (see Section \ref{sec:model:info} for details).
Notice that in the PU-proposal market, PUs propose offers and SUs respond by rejecting or accepting offers.
Under complete information,
each PU knows the whole network information, and can precisely predict the SUs' preferences.
Thus, SUs cannot mis-represent their preferences, and have to respond in the \emph{truthful} manner.
That is, {each SU will always accept the best offer that brings him the maximum positive utility.}
Under incomplete information, however, each PU knows only its local network information, and cannot predict the SUs' preferences with other PUs.
Thus, SUs can potentially mis-represent their preferences, and respond in a \emph{non-truthful} manner.
That is, {each SU may reject the best offer that brings him the maximum positive utility, as long as such a mis-representation can bring him a better utility.}
%\footnote{This can be shown by Example \ref{examp:2}. Suppose SU $n_1$ misrepresents its preferences by $Q(n_1) = \{m_2, m_3\}$. Then SU $n_1$ will reject the PU $m_1$'s offer, even when PU $m_1$'s offer is the best offer to him (for example, when PU $m_1$'s is the single user proposing to him).}}

\subsection{Equilibrium under Complete Information}

%We first consider the complete information scenario, where each PU knows the whole network information.
%%In this case, SUs cannot mis-represent their preferences.
%%We only need to consider the   collective interactions between the users.
%%Notice that in the PU-proposal market,
%%% under complete information,
%%PUs propose offers and SUs respond by rejecting or accepting offers.
\revv{Under complete information, each SU has to respond in a \emph{truthful} manner, and accept the best offer that brings him the maximum positive utility.
In this sense, we can view the SUs' (truthful) responses as natural reactions.
Therefore, the matching problem can be viewed as an \emph{$M$-player game with complete information}, where game players are PUs, whose objectives are to maximize their own utilities.}
%whose strategies are to decide which SUs to propose and how much utilities to offer to the proposed SUs}.
%\footnotesc{Each SU will respond to PUs' strategies truthfully, and PUs can take SUs' responses into account when making their own decisions. \revj{Thus, we do not need to consider SUs as strategic players.}}
%\revh{Similar to  Section \ref{sec:model}.C,} \revh{we call the equilibrium of this game as \emph{{\POEQ}}.}
%\comv{Since we have complete information here, is there still an issue of "truthfulness"? When we talk about truthfulness, seems that SUs still have a choice.}
By (\ref{eq:opt_pu}), each PU's maximal utility strictly decreases with the utility of the matched SU. Thus, given all other PUs' offers (strategies), the best strategy for a particular PU $\mu_n$ (i.e., the PU matched to SU $n$) is to offer the \emph{lowest acceptable} utility to its matched SU $n$, i.e., ${\delta}_n = \underline{\delta}_n$.
\revh{This essentially lead to the Pareto-optimal equilibrium for PUs ({\POEQ}).}
Next we  \revh{show how to compute} the {\POEQ} in the complete information scenario.
\revh{Notice that when $\mu(\cdot)$ is given, the {\POEQ} can be computed by jointly solving (\ref{eq:funcset}), which is complicated as mentioned previously.}
\rev{Moreover, an exhaustive search of all feasible $\mu(\cdot)$ requires going through} a total of $M!$ \revj{possibilities}, which has an exponential-time complexity.
This makes the solving process mathematically intractable.

To this end, we introduce a ``\emph{Generalized Deferred Acceptance}'' (G-DAC) algorithm in Algorithm \ref{alg:G-DAC}, which converges to \revj{the} {\POEQ} in polynomial-time complexity.
Specifically, in the G-DAC algorithm, each PU $m$ maintains a vector $\Delta^m=(\delta_1^m, ..., \delta_N^m)$, which records the utilities it is willing to offer to each SU.
The basic idea of the G-DAC algorithm is:  {(i) in each round, each PU $m$ proposes to the best SU based on $\Delta^m$, and each SU $n$ accepts the best proposal and rejects others; and (ii) if PU $m$ is rejected by SU $n$ in a round, it increases the offer $\delta_n$ by a \rev{small positive value} $\epsilon$ in the following rounds.}
%\begin{table} [h]
%\hspace{-5mm}
%\small
%\begin{tabular}{m{0.5\textwidth}}
%  \textbf{}
%\\
%\hline
%\textbf{Initial Stage}: \\
%For each PU $m$, set $\delta_n^m = 0,\forall n=1,...,N$;\\
%\hline
%\textbf{Iterative Stage}: \\
%For each PU $m$, propose to the best SU $n^*=\arg \max_{n} (f_n^m(\delta_n^m)\geq 0)$; \\
%For each SU $n$, reject all PUs but the one offering him highest utility; \\
%For each PU $m$, update $\delta_n^m = \delta_n^m + \epsilon $ if rejected by SU $n$; \\
%\hline
%\textbf{Converging Stage}: \\
%All $\delta_n^m$ do not change.\\
%\hline
%\end{tabular}
%\end{table}
%

\begin{algorithm}[h]
\small
\begin{algorithmic}[1]
\State \textbf{Initialization}
\State \hspace{1mm} Each PU $m\in\M$ \textbf{set} $\delta_n^m = 0,\forall n=1,...,N$;
\State \textbf{do} \{
\State  \hspace{1mm} Each PU $m$ \textbf{proposes} to $n$ if $f_n^m(\delta_n^m)\geq \max_{k}\{ f_k^m(\delta_k^m), 0 \}$;
\State \hspace{1mm} Each SU $n$ \textbf{accepts} the PU offering him highest utility;
\State \hspace{1mm} Each PU $m$ \textbf{updates} $\delta_n^m = \delta_n^m + \epsilon $ if rejected by SU $n$;~~~~\}
%\State \textbf{Converging Stage:}
%\State \hspace{1mm} All $\delta_n^m$ do not change.
\State \textbf{while} (at least one $\delta_n^m$ is changed)
\end{algorithmic}
\caption{G-DAC}\label{alg:G-DAC}
\end{algorithm}

\revv{The G-DAC algorithm generalizes the DAP algorithm \cite{gale1962college} in the following aspect.
In the DAP algorithm, when a PU is rejected by an SU, it will propose to another SU in   later rounds (as the   SU who has rejected its offer  will never accept its offer in the future); while in the G-DAC algorithm, the PU will continue to  increase the utility to the SU (attempting to attract the SU),
until the offer is accepted by the SU or the PU can achieve a higher utility by proposing an offer to another SU.}\footnote{Intuitively, the G-DAC algorithm is equivalent to an \revj{\emph{English-Auction}} \cite{krishna2009auction} with SUs as auctioneers and PUs as bidders. The \revj{auction works as follows:} {(i) each SU asks a ``price'' (the required utility), and gradually increases the price if multiple PUs bid to him, and (ii) each PU \rev{submits} bid to \rev{the best} SU, according to \rev{the maximum utilities that it can achieve from} SUs at the current price \rev{level} of all SUs.}}
%\revv{Note that this change is necessary for the two-sided market with transferrable utility, as the increase of utility (to be offered to an SU) essentially reflects the different utility transfers between the PU and the SU.}
\revv{Hence, it brings an additional degree of freedom for PUs making their decisions. Namely, each PU decides not only the target SU that it is going to propose to, but also the utility that it is going to transfer to the SU.}
%\comv{Why this change is needed/important, and how such change will complicate the analysis? These are important issues and should be included in the main texts. }
%\revj{The utilities of SUs increase, and the utilities of PUs decrease.}

%\rev{Overall, this is} an increasing-offer algorithm for SUs \rev{and a decreasing-offer algorithm for PUs.}

%The convergence of G-DAC is shown as follows.

\begin{lemma}\label{lemma:conv}
The G-DAC algorithm converges to the \revh{Pareto-optimal equilibrium} for PUs ({\POEQ}), if the stepsize $\epsilon$ is small engouth.
%$\epsilon<\bar{\epsilon}$, where $\bar{\epsilon}$ is a small positive value.
%\comv{can we characterize this threshold explicitly? Answer: No.}
\end{lemma}

To prove the above lemma, we can first show that the convergence state of G-DAC satisfies the conditions in Theorem \ref{lemma:necsuff1}, and is therefore an equilibrium.
Then we can prove the optimality by showing that the G-DAC will stop at the first equilibrium it achieves, which is exactly the Pareto-optimal equilibrium for PUs.

Next we discuss   the communication overhead and computational complexity of the G-DAC algorithm. In each round, each PU $m$ needs to send one request message to a particular SU, and each SU $n$ needs to send back one acceptance message to a particular PU. Thus, the communication overhead is linear in the number of PU-SU pairs (and hence is low). In terms of computation, each PU $m$ needs to solve the optimization problem (\ref{eq:opt_pu}) at most once with complexity $\rho $ in each round (except the first round where the PU needs to solve the optimization problem (\ref{eq:opt_pu}) $N$ times with complexity $N \rho $), since at most one price in the  vector $\Delta^m=(\delta_1^m, ..., \delta_N^m)$ is changed in each round. Furthermore,
each PU needs to find the maximal utility among $N$ utilities,
% \revyy{(with the worst case of $N$ units of computational burden)},
and each SU needs to find the best offer among at most $M$ proposals.
% \revyy{(with the worst case of $M$ units of computational burden)}.
Thus, the total computational complexity is  $\mathcal{O}(T\cdot (M   \rho + M  {N} + N   M))$,
%\comv{Lin: can we write the complexity in such an $O(\cdot)$ notation?}
where $T = \frac{\sum_{n\in\N} \underline{\delta}_n^*}{\epsilon} $ is the maximum possible iterations, and $\rho $ is the complexity  of  solving problem (\ref{eq:opt_pu}).
%where $\rho^{UTF}$ is the computational burden for solving optimization (\ref{eq:funcset}), $\rho^{N} $ and $ \rho^M$ are the computational burden for finding the maximal number from a set of size $N$ and $M$, respectively.
%\comv{what is the complexity of solving (11)?
%Without this, the discussions here are not complete. }

%
%\begin{figure*}
%\hspace{-1mm}
%  \begin{minipage}[t]{0.5\linewidth}
%    \includegraphics[scale=.45]{simu1-new}
%    \caption{Equilibria, UTF, and GS-UTF with 1 PU and 1 SU.} \label{f-simu1}
%    \end{minipage}
%  \begin{minipage}[t]{0.005\linewidth}~
%  \end{minipage}
%  \begin{minipage}[t]{0.5\linewidth}
%    \includegraphics[scale=.45]{simu2-new}
%    \caption{Equilibria, UTF, and GS-UTF with 1 PU and 2 SU.} \label{f-simu2}
%  \end{minipage}
%\end{figure*}

\subsection{Equilibrium under Incomplete Information}

Under incomplete information, each SU can respond in a \emph{non-truthful} manner, and may reject the best offer that brings him the maximum positive utility, as long as such a mis-representation can bring a better utility for the SU.
This can be seen from the following example.
\begin{example}
Suppose PU $\mu_n$ proposes the \emph{lowest acceptable} utility to its matched SU $n$ (as in the complete information scenario). The SU $n$ can simply reject this offer (which is its best offer for SU $n$  according to the definition of the {lowest acceptable} utility), such that PU $\mu_n$ increases the utility offered to SU $n$. \hfill $  \square$
\end{example}

%\begin{table} [h]
%\hspace{-5mm}
%\small
%\begin{tabular}{m{0.51\textwidth}}
%  \textbf{Generalized Reversed Deferred Acceptance -- G-RDAC}
%\\
%\hline
%\textbf{Initial Stage}: \\
%For each SU $n$, set $\pi_n^m = 0,\forall m=1,...,M$;\\
%\hline
%\textbf{Iterative Stage}: \\
%For each SU $n$, propose to the best PU $m^*=\arg \max_{n} (g_n^m(\pi_n^m)\geq 0)$; \\
%For each PU $m$, reject all SUs but the one offering him highest utility; \\
%For each SU $n$, update $\pi_n^m = \pi_n^m + \epsilon $ if rejected by PU $m$; \\
%\hline
%\textbf{Converging Stage}: \\
%All $\pi_n^m$ do not change.\\
%\hline
%\end{tabular}
%\end{table}
%
%

%
%\begin{algorithm}[h]
%\small
%\begin{algorithmic}[1]
%\State \textbf{Initial Stage:}
%\State \hspace{2mm} Each PU $m$ \textbf{set} $\Pi_m = 0$;
%\State \textbf{Iterative Stage:}
%\State \hspace{2mm} Each PU $m$ calculate the offer for every SU based on $\Pi_m$;
%\State \hspace{2mm} Each SU $n$ \textbf{propose} to $m$ if $g_n^m(\Pi_m)\geq \max_k\{g_n^k(\Pi_k) , 0\}$;
%\State \hspace{2mm} Each PU $m$ \textbf{update} $\Pi_n^m = \Pi_n^m + \epsilon $ if receiving multiple proposals;
%\State \textbf{Converging Stage:}
%\State \hspace{2mm} All $\pi_n^m$ do not change.
%\end{algorithmic}
%\caption{\label{alg:G-DAC}Generalized Reversed Deferred Acceptance -- G-RDAC}
%\end{algorithm}

\revv{By the above example, we can see that different mis-representation behaviors of SUs (e.g., determining when to accept a PU's offer, and when to reject) may lead to different market equilibria.
Unfortunately, due to the continuity of utility, there are infinite number of mis-representation behaviors of SUs.
Thus, it is impossible to characterize all possible mis-representation behaviors of SUs.
To this end, we will focus on characterizing a \emph{robust} equilibrium for PUs ({\RBEQ}), which provides every PU a guaranteed~utility under any possible misrepresentations of SUs.

Similarly, the robust equilibrium for PUs (\RBEQ) can be achieved by a ``\emph{Generalized Reversed Deferred Acceptance}'' (G-RDAC)  algorithm in Algorithm \ref{alg:G-RDAC}, which is same as the G-DAC Algorithm \ref{alg:G-DAC} except that it reverses the roles of PUs and SUs in proposing offers.\footnote{Similarly, \rev{we can show that} the G-RDAC algorithm is equivalent to an English-auction with PUs as auctioneers and SUs as bidders.}
% : \emph{(i) each PU asks a ``price'' (its reserved utility) and gradually increases its price if multiple SUs proposing him, and (ii) each SU proposes bid on a particular PU based on its achieved utilities on different PUs at current price set.} Thus, it is an increasing-offer algorithm for PUs, while decreasing for SUs.
Since the G-DAC algorithm and G-RDAC algorithm are symmetric, \rev{we can easily show that} the G-RDAC algorithm \revj{converges} to the Pareto-optimal equilibrium for SUs, which is also the worst-case equilibrium for PUs.
%The convergence of the G-RDAC can be proved in a similar way as that of the G-DAC given in Lemma \ref{lemma:conv} for the G-DAC.
%The details are skipped due to the page limit.
%Due to the conflict interests of PUs and SUs (regarding equilibrium), the Pareto-optimal equilibrium for SUs (achieved in G-RDAC) is also the worst equilibrium for PUs.
Therefore, every PU is guaranteed to achieve a utility no worse than that in this worst-case equilibrium, under any possible misrepresentations of SUs.
In this sense, we refer to this worst-case equilibrium  for PUs as the robust equilibrium for PUs.}

\begin{algorithm}[h]
\small
\begin{algorithmic}[1]
\State \textbf{Initialization}
\State \hspace{2mm} Each SU $n$ \textbf{sets} $\pi_n^m = 0,\forall m=1,...,M$;
\State \textbf{do} \{
\State \hspace{2mm} Each SU $n$ \textbf{proposes} $m$ if $g_n^m(\pi_n^m)\geq \max_k\{g_n^k(\pi_n^k) , 0\}$;
\State \hspace{2mm} Each PU $m$ \textbf{accepts} the SU offering him highest utility;
\State \hspace{2mm} Each SU $n$ \textbf{updates} $\pi_n^m = \pi_n^m + \epsilon $ if rejected by PU $m$; ~~~\}
%\State \textbf{Converging Stage:}
%\State \hspace{2mm} All $\pi_n^m$ do not change.
\State \textbf{while} (at least one $\pi_n^m$ is changed)
\end{algorithmic}
\caption{G-RDAC}\label{alg:G-RDAC}
\end{algorithm}

\subsection{Equilibrium under Arbitrary Information}

\revv{It is important to note that the {\RBEQ} guarantees every PU's lowest achievable utility among all equilibria, while the {\POEQ} ensures every PU's highest achievable utility among all equilibria.
This implies that under any possible equilibrium (in any  information scenario), every PU's utility must be bounded by its utilities under the  {\RBEQ} and  the {\POEQ}.
Our numerical studies in the next section further show that in a typical market where the number of PUs and SUs are different, the   gap between the {\RBEQ} and the {\POEQ} is quite small. Hence, the {\RBEQ} or {\POEQ} can be viewed as an effective approximation to understand all equilibria.}

\begin{figure*}[tt]
\vspace{-3mm}
\centering
\includegraphics[width=0.4\textwidth]{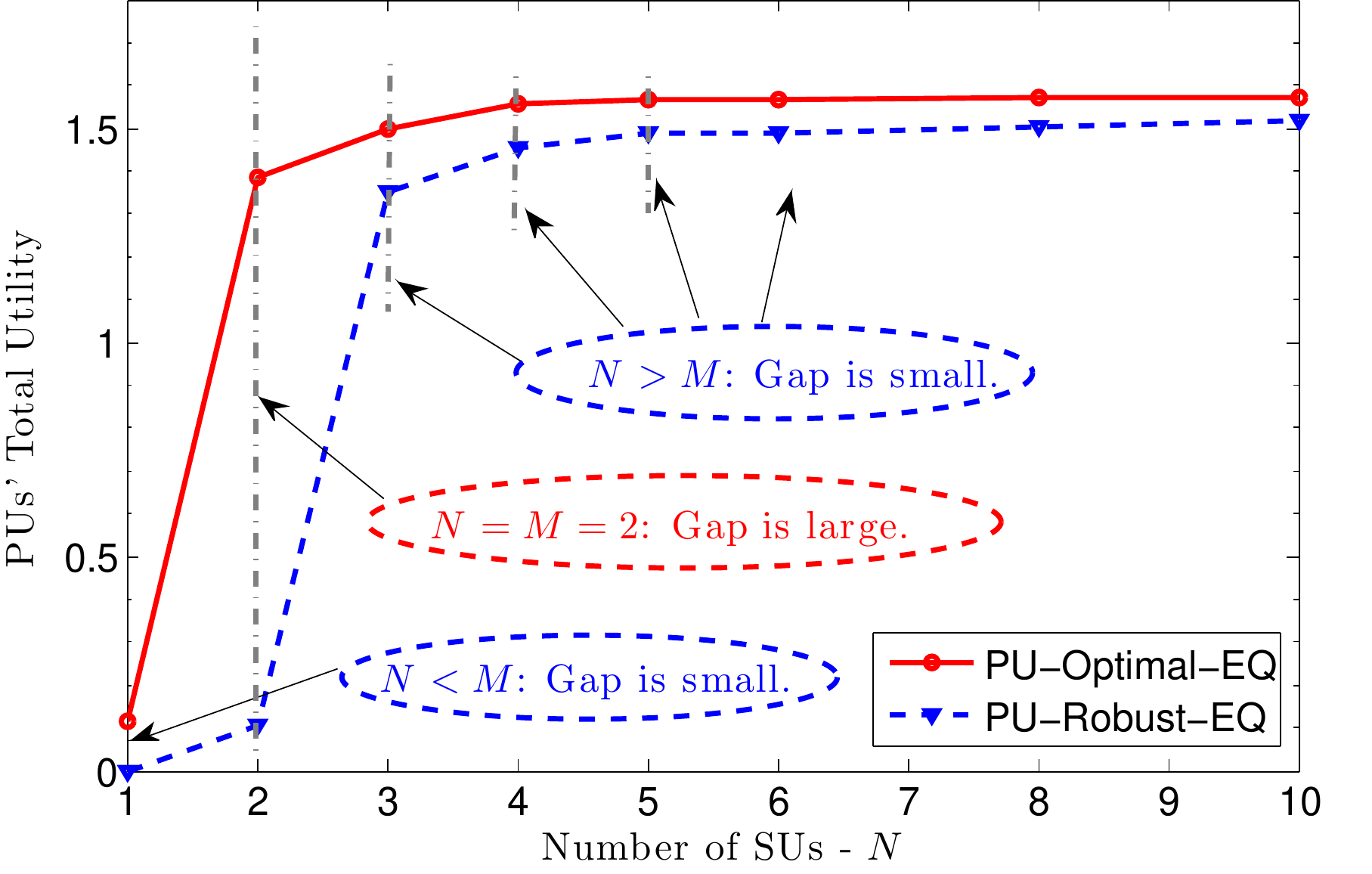}
~~~~~~~~~~\includegraphics[width=0.4\textwidth]{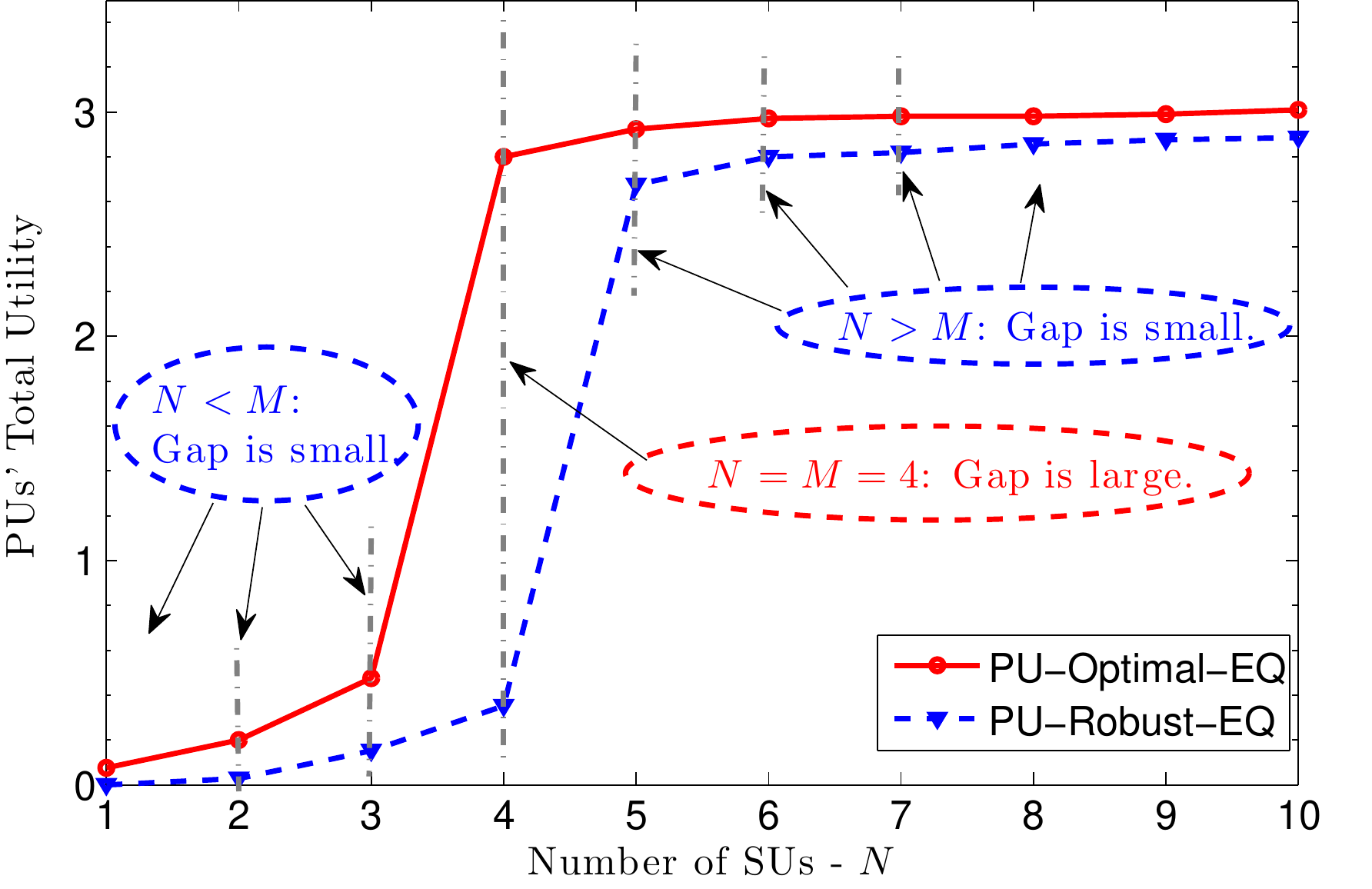}
\caption{Total PUs' utility vs Number of SUs $N$. The number of PUs is $M=2$ in the left figure, and  $M=4$ in the right figure. The PUs' utility gap between {\POEQ} and {\RBEQ} is less than $10\%$ when $M\neq N$.} \label{f-simu4}
\vspace{-4mm}
\end{figure*}

\section{Simulation Results}\label{sec:simu}

To highlight the performance gain achieved from cooperative spectrum sharing, we focus on such a network scenario in simulations, where PUs' direct channel gains are \emph{much smaller} than  SUs' relay channel gains. This may happen when PUs' direct links are highly attenuated due to shadow fading or obstacles. This is very intuitive: only when an SU can provide a large enough benefit to a PU, the PU has the incentive to cooperating with the SU, which is a prerequisite of the cooperative spectrum sharing studied in this paper.

We distribute PUs' and SUs' transceiver pairs in a square  area of size $1500 \times 1500 $m$^2$, and set the distance of each PU's transceiver pair close to $1000$m and the distance of each SU's transceiver pair close to $400$m \revv{(which are the typical transmission ranges between mobile phones and cellular base stations).}
%We further set each SU's transmitter close to the middle of a PU's transceiver pair.
\revmm{Unless otherwise stated, the following default network setting will be used: (i) $C_n = T_m =1 $, (ii) $\sigma^2 = -105$dBm, and (iii) the channel gains are chosen based on the free-space path loss model in \cite{goldsmith2005wireless}.}
%\revv{These parameters are chosen according to the settings of a typical cellular network (see \cite{goldsmith2005wireless} for details)}.
%\footnote{\revmm{It is important to note that our purpose in the current simulations is to provide some intuitive illustrations and meaningful insights into the equilibrium
%in these general two-sided models.
%Simulations using realistic parameters and network environments (such as cellular models based on 3GPP SCM \cite{ch-1} and WINNER 2 \cite{ch-2}) are also very important for motivating the proposed problem/solution formulation as well as the performance evaluations.
%In our future work, we will construct the experimental simulation environments and demo systems based on more detailed wireless models (such as the widely-used WINNER 2 model \cite{ch-2}).}}

%\subsection{Performances Evaluation}
We evaluate the PUs' performance under different market equilibria (illustrations of  equilibria can be found in \cite{techrpt}). 
Note that the SUs' performance can be estimated directly from the conflicting interests between PUs and SUs, that is, \rev{an equilibrium that is better for all PUs must be worse for all SUs.}
\revmm{In the following, each result is the average of 1000 simulations with randomly generated network topologies.}

\revv{Figure \ref{f-simu4} illusrates the PUs' total utilities  \texttt{vs} the number of SUs $N$ under different market equilibria, where the number of PUs is $M=2$ in the left figure, and $M=4$ in the right figure.
%In this simulation, PUs' direct channel gains are randomly drawn from $[-100, -120]$dBm, and relay channel gains and SUs' channel gains are randomly drawn from $[-85, -95]$dBm.
The red curve denotes the total PUs' utility under  the \POEQ, and the blue dash curve denotes the total PUs' utility under   the \RBEQ. Obviously, the total PUs' utility under any other equilibrium is bounded by these two curves.~~~~~~~~~~~~~~~

From Figure \ref{f-simu4}, we can see that the PUs' total utility under equilibrium increases with the number of SUs $N$ (as the competition among SUs increases with $N$), and decreases with the number of PUs $M$ (as the competition among PUs increases with $M$).
Moreover, when $M<N$, PUs can extract most of the generated  utility (as the competition among SUs is more intense), and when $M>N$, SUs can extract most of the generated utility (as the competition among PUs is more intense).
We further have the following observations.

(1) In the general case where the numbers of PUs and SUs are different (i.e., $M\neq N$),
the utility gap between {\POEQ} and {\RBEQ} is  quite small (e.g., less than $10\%$ in both figures). This implies that the information scenario (i.e., the amount of network information that PUs know) has a   small impact on the market equilibrium.

(2) In the special case where the numbers of PUs and SUs are identical (i.e., $M=N$), the utility gap between {\POEQ} and {\RBEQ} is large.
This implies that the information scenario has a significant impact on the market equilibrium.
%\item
%The PUs' performances increases with the number of SUs $N$ as the competition among SUs increases with $N$, and decreases with the number of PUs $M$ as the competition among PUs increases with $M$;
%\item
%When the competition levels among PUs and SUs are unbalanced ($N \neq M$), the PUs' performances gap between PU-Dominant-EQ and PU-Pessimistic-EQ (or PU-Pessimistic-EQ-GS) is very small, and this implies that \textbf{the impact of
%market competition is larger than the network information};
%When $N=M$, this gap is relatively large.

Intuitively, when $N\neq M$, the imbalance of competition among PUs and among SUs reduces the effect of information incompleteness.
The reason is as follows.
If the number of SUs is larger than that of PUs ($N>M$),
the intensified competition among SUs pushes SUs to reveal their preferences close to the true values.
If the number of PUs is larger than that of SUs ($M>N$),
the intensified competition among PUs already drives most of the generated welfare to the SU-side even in the worst-case equilibrium for SUs, and thus the SUs' performance gain in other equilibria is small.
When $M=N$,  however, \revj{none of the two sides has a dominated competition level in the market, thus the effect of information incompleteness becomes significant.}}

%!TEX root = CoopSpecShar_main.tex
%SourceDoc CoopSpecShar_main.tex

\section{Conclusion}\label{sec:con}

In this paper, we study the cooperative spectrum sharing between {multiple PUs and multiple SUs using the matching theory}.
We formulate the problem as a two-sided matching market, and  characterize the market equilibrium systematically.
We further study which equilibrium will actually emerge in both complete and incomplete information scenarios.
%We further study the impacts of market structure and network information on the equilibrium.
%Our work reveals that two-sided matching market can be a useful tool to analyze large networks where two types of players interact with each other independently.
%The results suggest that in an unbalanced market, the emerging equilibria are very similar no matter how much information players know.
Our study can facilitate the design of large networks, in which players can be split into two different types, interacting with each other.
The interaction between players is not restricted within the cooperative spectrum sharing studied in this work, but can be quite general (e.g., spectrum trading).
%The matching equilibrium characterized in this study  provides a good prediction for the \emph{network stable state} resulting from the collective interactions among players.
%Furthermore, the deferred acceptance algorithms enable distributedly converge to the stable state.
%\comv{what about future work?}
%It is worthy noting that the prediction of a large network's stable state is nontrivial in practice, and can provide instruction in different objectives, e.g., network optimizating.
\revv{There are some possible directions to extend
the results in this paper. An interesting direction is to consider a more general cooperative spectrum sharing protocol where each PU can cooperate with multiple SUs to relay its traffic, and each SU can cooperates with multiple PUs to access their spectrum.
In this case, the resulting matching is no longer a one-to-one correspondence between PUs and SUs, but a multiple-to-multiple correspondence between PUs and SUs.}

%!TEX root = CoopSpecShar_main.tex
%SourceDoc CoopSpecShar_main.tex

%!TEX root = CoopSpecShar_main.tex
%SourceDoc CoopSpecShar_main.tex

\begin{biography}[{\includegraphics[width=1in,clip,keepaspectratio]{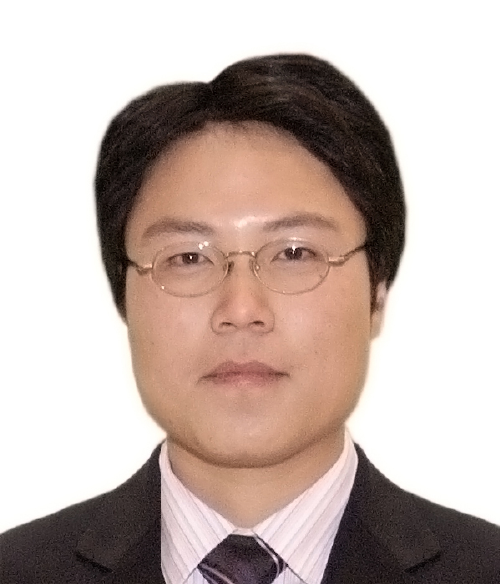}}]{Lin Gao} 
(S'08-M'10) is an Associate Professor in the School of Electronic and Information Engineering, Harbin Institute of Technology (HIT) Shenzhen Graduate School.
He received the M.S. and Ph.D. degrees in Electronic Engineering from Shanghai Jiao Tong University (China) in 2006 and 2010, respectively.
He worked as a Postdoc Research Fellow at the Chinese University of Hong Kong from 2010 to 2015.
His research interests are in the area of network economics and games, with applications in wireless communications, networks, and internet of things.
\end{biography}

\begin{biography}[{\includegraphics[width=1in,clip,keepaspectratio]{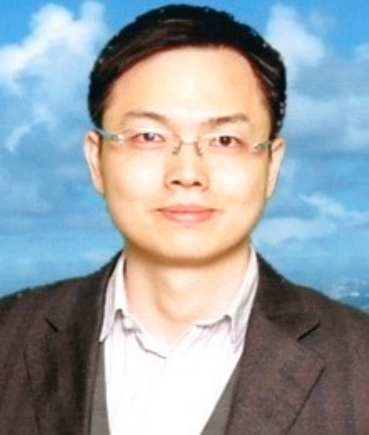}}]{Lingjie Duan}
(S'09-M'12) is an Assistant Professor with Engineering Systems and Design, Singapore University of Technology and Design (SUTD), Singapore. He received the Ph.D. degree from the Chinese University of Hong Kong, Hong Kong, in 2012. In 2011, he was a Visiting Scholar University of California at Berkeley, CA, USA. He is an Editor of IEEE Communications Surveys and Tutorials, and is a SWAT member in the Editorial Board of IEEE Transactions on Vehicular Technology. He is also a Guest Editor of IEEE Wireless Communications magazine for a special issue about green networking and computing for 5G. His research interests include network economics and game theory, cognitive communications and cooperative networking, energy harvesting wireless communications, and network security. Recently, he served as the Program Co-Chair of the IEEE INFOCOM 2014 GCCCN Workshop, ICCS 2014 special session on Economic Theory and Communication Networks, the Wireless Communication Systems Symposium of the IEEE ICCC 2015, the GCNC Symposium of the IEEE ICNC 2016, and the IEEE INFOCOM 2016 GSNC Workshop. He also served as a technical program committee (TPC) member of many leading conferences in communications and networking (e.g., ACM MobiHoc, IEEE SECON, ICC, GLOBECOM, WCNC, and NetEcon). He was the recipient of the 10th IEEE ComSoc Asia-Pacific Outstanding Young Researcher Award in 2015, the Hong Kong Young Scientist Award (Finalist in Engineering Science track) in 2014, and the CUHK Global Scholarship for Research Excellence in 2011.
\end{biography}

\begin{biography}[{\includegraphics[width=1in,clip,keepaspectratio]{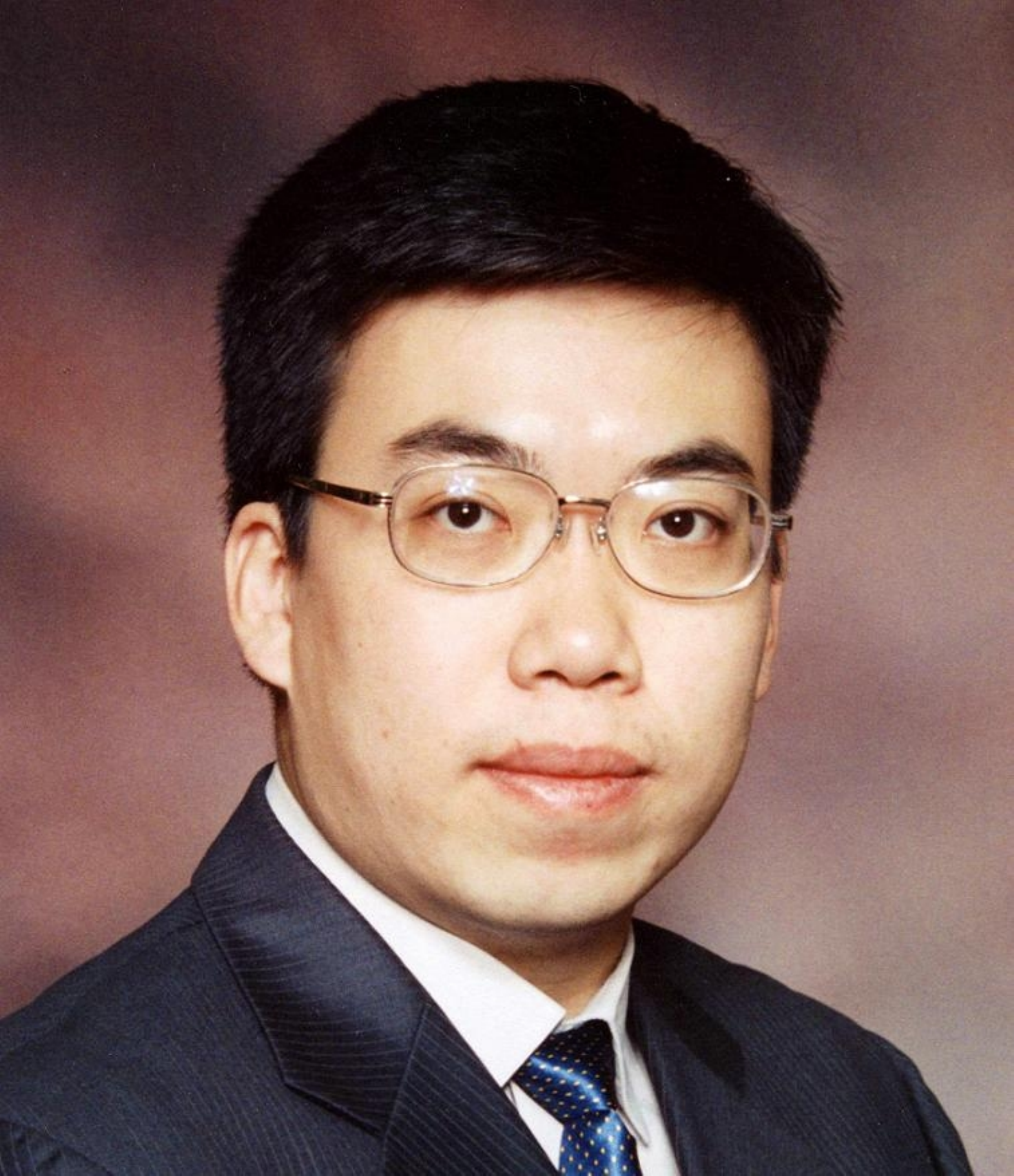}}]{Jianwei Huang}
(S'01-M'06-SM'11-F'16) is an Associate Professor and Director of the Network Communications and Economics Lab (ncel.ie.cuhk.edu.hk), in the Department of Information Engineering at the Chinese University of Hong Kong. He received the Ph.D. degree from Northwestern University in 2005, and worked as a Postdoc Research Associate in Princeton University during 2005-2007. He is the co-recipient of 8 international Best Paper Awards, including IEEE Marconi Prize Paper Award in Wireless Communications in 2011. He has co-authored four books: "Wireless Network Pricing," "Monotonic Optimization in Communication and Networking Systems,"  "Cognitive Mobile Virtual Network Operator Games,¡± and "Social Cognitive Radio Networks". He has served as an Associate Editor of IEEE Transactions on Cognitive Communications and Networking, IEEE Transactions on Wireless Communications, and IEEE Journal on Selected Areas in Communications - Cognitive Radio Series. He is the Vice Chair of IEEE ComSoc Cognitive Network Technical Committee and the Past Chair of IEEE ComSoc Multimedia Communications Technical Committee. He is a Fellow of IEEE and a Distinguished Lecturer of IEEE Communications Society.
\end{biography}

\newpage
\appendix
%!TEX root = Section5-General.tex
%SourceDoc Section5-General.tex

Outline of this appendix:
\begin{itemize}
  \item A. Strongly Incomplete Information 
  \item B. Illustrations of Equilibria in Simulations
  \item C. Implementation Issues
  \item D. Equilibrium Analysis for the Simplified Model
  \item E. Proof for Theorem \ref{lemma:necsuff1}
  \item F. Proof for Lemma \ref{lemma:lattice}
  \item G. Proof for Theorem \ref{lemma:opt}
\end{itemize}

\subsection*{A.  Strongly Incomplete Information}\label{sec:strongly}

In the main manuscript, we have studied the complete information and incomplete information scenarios. 
Now we consider the strongly incomplete information scenario, where each PU only knows its own channel and relay channel information, but not SUs' channel information.
\revj{In this case, each PU does not have enough information to solve the optimization problem (\ref{eq:opt_pu}), even if it knows the reservation utility of its matched SU.}
%does not know each SU's lowest acceptable utility, or any other network information (e.g., the SU's direct channel gain and  sensitivity to power consumption\com{ ratio}) in solving optimization optimal (\ref{eq:opt_pu}).
Thus, we need to consider not only the mis-representations of SUs, but also \rev{the proposing rule with which PUs can calculate the desired offers to SUs.~~~~~~~~~~}
%the proposing rule of PUs (since the proposing rule given in (\ref{eq:opt_pu}) is not applicable due to the lack of network information).
%

To this end, \rev{we will first design a \emph{proposing rule} for PUs based on their ``guesses'' about SUs' information}. Then we will study the incentive for SUs to truthfully represent their preferences under the new proposing~rule, and \revj{characterize the corresponding} equilibrium.

The basic idea \revj{of the guess-based proposing rule} is \rev{as follows}. First, each PU $m$ {guesses} the necessary information (called type) about every SU, and proposes an offer aiming at \emph{extracting the entire cooperation gain from an SU with the guessed information}.\footnotesc{\revmm{Note that we can neither prove the optimality of this proposing rule, nor find another rule always better than this one. Nevertheless, through simulations we can see that the PUs' performances with this proposing rule are very closed to those with optimal proposing rule (\ref{eq:opt_pu}) under weakly incomplete information.}} If it is rejected by an SU, the PU updates its guess about this SU, and propose a new offer (aiming at extracting the entire cooperation gain from the SU with the updated guess).

%For simplicity and without loss of generality, we assume   each SU $n$ has the same (and public) direct channel gain on all bands, i.e., $G_{n(m)}= G_n, \forall m \in \mathcal{M}$. Thus, the only uncertain information for PUs is SUs' power consumption sensitivity.\footnotesc{Without this assumption, the PU has to guess a two-dimension information, which greatly increases the complexity, while not introduces any new insight.}
To \revh{capture} the \revh{key} information \rev{that PU $m$ wants to know about} SUs, we introduce the concept of \emph{type} for each SU $m$, denoted by  $H_n^m$ and defined as
$$\textstyle
H_n^m = 2\cdot \frac{R_{n(m)}-C_n}{C_n  T_m}.
$$
%\rev{which we refer to as} the \emph{type} of SU $n$ for PU $m$.
%Notice that $H_n^m$  is obtained from $\Delta_{n}^m =0$.
Notice that $H_n^m$ captures all of the key information of SU $n$ (for PU $m$).
With the type $H_n^m$, the SU's utility defined in (\ref{eq:SU_ut}) can be rewritten as
\begin{equation}\label{eq:SU_ut-h}\textstyle
\Delta_{n}^m = (t_n^m \cdot H_n^m - p_n^m)\cdot A,
\end{equation}
where $A \triangleq \frac{C_n T_m}{2\cdot (T_m+   t_n^m)}$ is a constant.

By (\ref{eq:SU_ut-h}), we can easily find that to ask for a relay power  $p_n^m$ from SU $n$, PU $m$ has to reward an access time no smaller than $t_n^m \cdot H_n^m$ to SU $n$ (otherwise SU $n$ will get a negative utility, and thus will never accept the offer).
Obviously, PU $m$ is able to derive the optimal proposing rule to SU $n$ from (\ref{eq:opt_pu}) only if it knows the real type $H_n^m$.
In this scenario, however, PU $m$ cannot obtain the real type $H_n^m$ of SU $n$, since it does not know the direct channel gain $G_{n(m)}$ of SU $n$.

Let us denote $\widetilde{H}_n^m$ as the PU $m$'s \emph{guess} about the SU $n$'s real type $H_n^m$. With this guess $\widetilde{H}_n^m$, PU $m$ aims at extracting all the cooperation gain from SU $n$, i.e.,
\begin{equation}\label{eq:opt_pu_guess}
\begin{aligned}
\widetilde{\Pi}_n^{m*} &  =  \max_{\{p_n^{m }, t_n^{m }\}} \quad \Pi_n^{m} (p_n^{m }, t_n^{m })\\
&\textstyle s.t.\quad p_n^{m } = t_n^m \cdot \widetilde{H}_n^m.
\end{aligned}
\end{equation}

The constraint in (\ref{eq:opt_pu_guess}) shows that (i) if  $ H_n^m = \widetilde{H}_n^m$ (i.e., the guessed SU type is exactly same as the true SU type), then \rev{SU $n$ will achieve a zero utility from the PU $m$'s proposed offer given by (\ref{eq:opt_pu_guess})}, and (ii) if $ H_n^m > \widetilde{H}_n^m$ (or $ H_n^m < \widetilde{H}_n^m$), then SU $n$ can achieve a positive (or negative) utility from the PU $m$'s proposed offer.
Thus, under this guess-based proposing rule, SU $n$ will accept (or reject) the PU $m$'s offer only if its real type $H_n^m$ is larger (or smaller) than the PU $m$'s guess $\widetilde{H}_n^m$.
In addition, when accepting the PU's offer (i.e., $H_n^m > \widetilde{H}_n^m$), SU $n$'s actual utility is
\begin{equation}\label{eq:opt_su_guess}
\begin{aligned}
\textstyle
\widetilde{\Delta}_n^{m*} &\textstyle = \frac{ t_n^{m*} \cdot  (R_{n }-C_n) - p_n^{m*}\cdot \frac{ C_n  T_m }{2} }{T_m+   t_n^{m*}} \\
& = \textstyle \frac{ C_n  T_m }{2} \cdot \frac{t_n^{m*}}{T_m+   t_n^{m*}} \cdot (H_n^m - \widetilde{H}_n^m),
\end{aligned}
\end{equation}
where $ t_n^{m*}$ and $ p_n^{m*}$ are the optimal solution to (\ref{eq:opt_pu_guess}).

Note that  $\widetilde{\Pi}_n^{m*}$, $\widetilde{ \Delta}_n^{m*} $, $ t_n^{m*}$, and $ p_n^{m*}$ are all functions of $\widetilde{H}_n^m$.
Thus, (\ref{eq:opt_pu_guess}) and (\ref{eq:opt_su_guess}) define an implicit function between $\widetilde{\Pi}_n^{m*}$ and $\widetilde{\Delta}_n^{m*}$.
Similar to the Utility Transfer Function (UTF) defined in (\ref{eq:opt_pu}), we define a \emph{Guess-based Utility Transferring Function (GS-UTF)}, denoted by $\widetilde{\Pi}_n^{m*} \triangleq \widetilde{f}_n^m(\widetilde{\Delta}_n^{m*})$.
It is easy to check that $\widetilde{\Pi}_n^{m*}$ \revh{strictly} increases \rev{with $\widetilde{H}_n^m$}, and $\widetilde{\Delta}_n^{m*}$ \revh{strictly} decreases with $\widetilde{H}_n^m$.  \revj{Thus, $\widetilde{f}_n^m(\widetilde{\Delta}_n^{m*})$ is a \revj{strictly} decreasing function of $\widetilde{\Delta}_n^{m*}$.} Figure \ref{f-simu1} in   Section \ref{sec:simu} illustrates this guess-based utility transferring function.

%
%\begin{figure*}
%\hspace{-5mm}
%  \begin{minipage}[t]{0.33\linewidth}
%    \centering
%    \includegraphics[scale=.3]{simu1}
%    \caption{Equilibria, UTF, and GS-UTF with 1 PU and 1 SU.} \label{f-simu1}
%    \end{minipage}
%  \begin{minipage}[t]{0.005\linewidth}~
%  \end{minipage}
%  \begin{minipage}[t]{0.33\linewidth}
%    \centering
%    \includegraphics[scale=.3]{simu2}
%    \caption{Equilibria, UTF, and GS-UTF with 1 PU and 2 SU.} \label{f-simu2}
%  \end{minipage}
%  \begin{minipage}[t]{0.005\linewidth}~
%  \end{minipage}
%  \begin{minipage}[t]{0.33\linewidth}
%    \centering
%    \includegraphics[scale=.3]{simu4_ut}
%    \caption{PUs' total utility under different equilibria.} \label{f-simu4}
%  \end{minipage}
%\end{figure*}

Notice that \textbf{all results} based on  UTF in Section \ref{sec:main}.A-D can be \textbf{directly} applied to the model based on GS-UTF here. \revh{Specifically, the necessary and sufficient conditions for an equilibrium can be characterized by Theorem \ref{lemma:necsuff1} (by simply replacing all involved UTF functions into the GS-UTF functions). The optimality of equilibrium can be proved by a similar Lattice theorem in Lemma \ref{lemma:lattice}.}
\revv{Note that here the ``optimality'' refers to the optimal equilibrium \emph{restricted within the above guess-based proposing rule}, but not the globally optimal equilibrium in all possible proposing rules. As mentioned previously, it is challenging to characterize the globally optimal equilibrium.}
%, as we replace the UTF $f_n^m(\cdot)$ by GS-UTF $\widetilde{f}_n^m(\cdot)$.
%Besides, the equilibrium under UTF may not be the equilibrium under GS-UTF.
%\com{still not clear. what results are we referring to here? Given the lemma and theorem numbers.}

%However, in practice,
%Given the above proposing rule for PUs, we \revh{then} consider the \emph{{incentive issues}} for SUs.  \revh{Following the same approach as in Section \ref{sec:toy}.D or \ref{sec:main}.D}, \rev{we can show that} it is possible for SUs to misrepresent their preferences by lying {\revh{on}} their real types and thereby hiding their information on lowest acceptable utilities.
%Unfortunately, it is also  hard to characterize which preferences each SU will misrepresent under strongly incomplete information, and which equilibrium will emerge.

Similarly, we can obtain a \emph{lower-bound} equilibrium   \rev{(in terms of PUs' utilities)} by a ``\emph{Guess-based Generalized Reversed Deferred Acceptance}'' procedure (GSG-RDAC), which is the same as the G-RDAC except the utility transferring function. \revh{The details of the GSG-RDAS is skipped due to the space limit.}
%The basic idea of GSG-DAC is:  {(i) in every round, each PU $m$ proposes to the best SU based on $\Delta^m$, and each SU $n$ rejects all proposals but the highest; and (ii) if a PU $m$ is rejected by SU $n$ in any round, it increases the $\delta_n^m$ with  given number.} Essentially, G-DAC generalizes the DAC in the aspect that when rejected by an SU, a PU will turn to other SUs immediately in DAC, while in G-DAC, a PU will increase its utility to the SU, and compare its own utilities on this SU and other SUs.
We referred to the equilibrium resulting from the GSG-RDAC as \emph{PU-Pessimistic-EQ-GS}.
%, which is same as the SU-optimal equilibrium based on GS-UTF.
%, and therefore is a lower-bound (in terms of \revj{PUs' utilities}) of any equilibrum based on GS-UTF.
Obviously, PU-Pessimistic-EQ-GS guarantees the worst-case utility of every PU, with the above guess-based proposing rule, under all possible mis-representations of SUs.

\subsection*{B. Illustrations of Equilibria in Simulations}

Now we illustrate the equilibria (i.e., utility division and relay assignment) in simulations. 

\subsubsection*{(1) Utility Division under Equilibrium}

\begin{figure*}[tt]
\centering
\includegraphics[scale=.4]{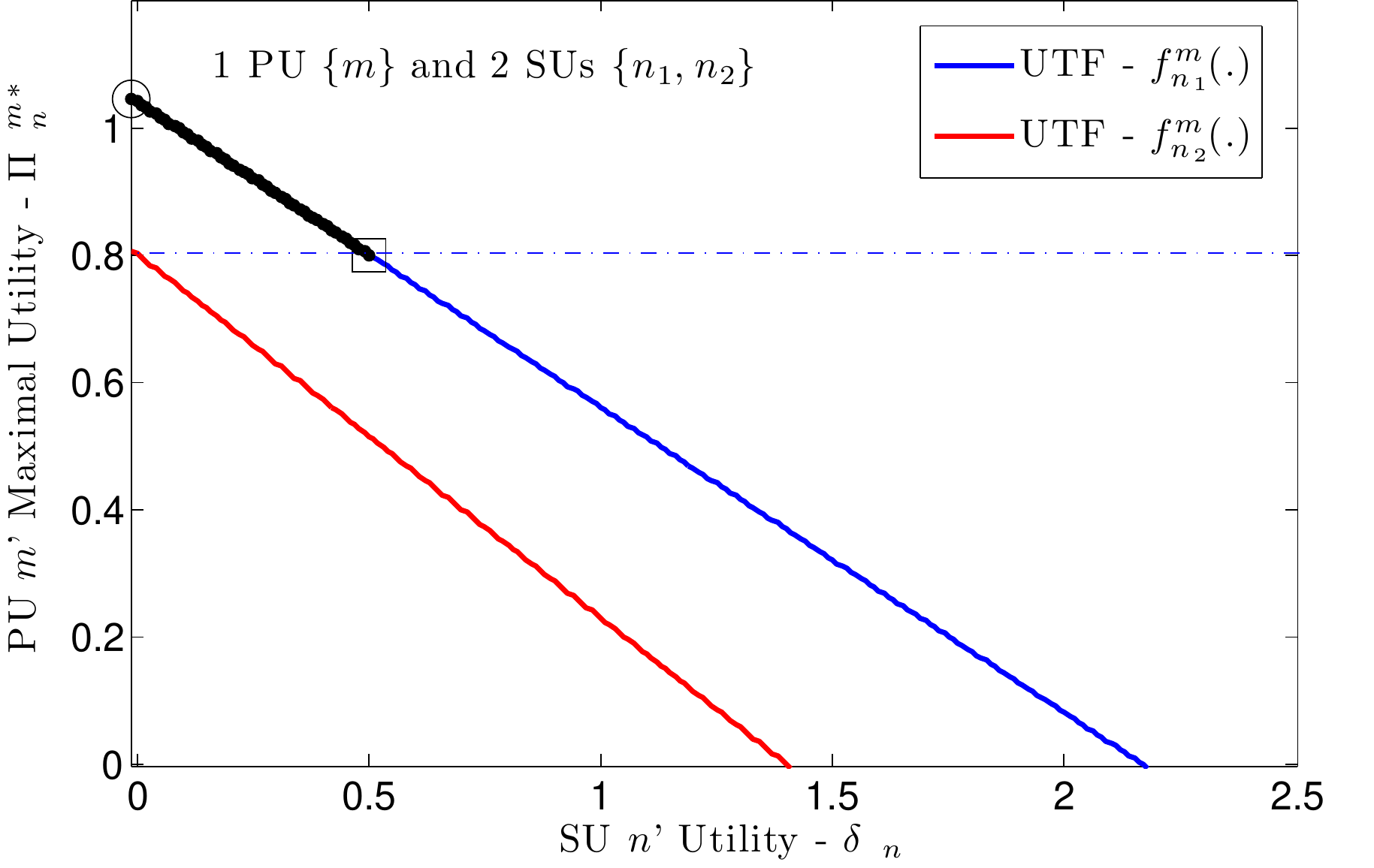} %{simu1-new-simple}
~~~~~~\includegraphics[scale=.4]{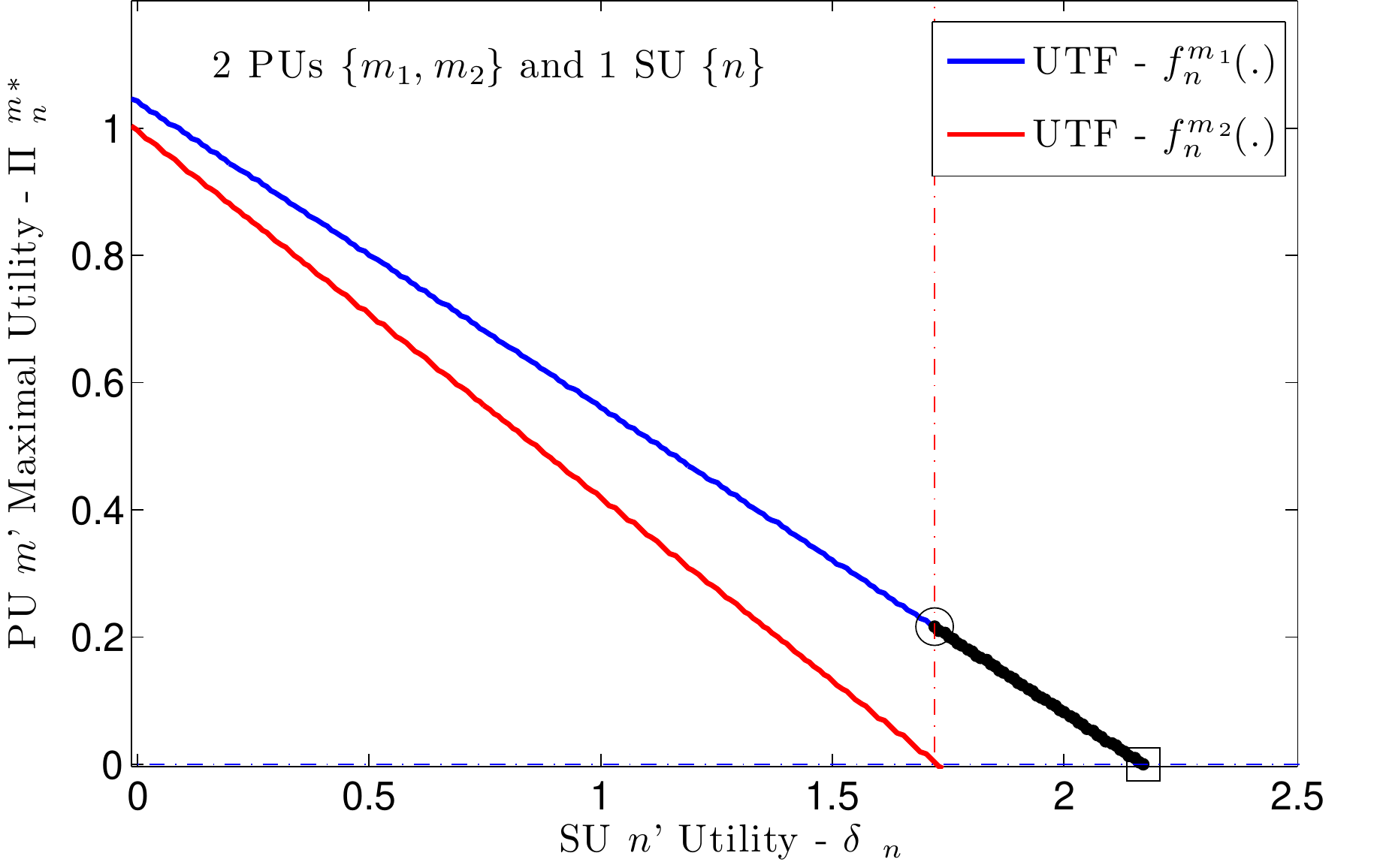} %{simu1-new-simple}
%    \caption{Illustration of UTF, GS-UTF, and related equilibria in a network with 1 PU and 1 SU.} \label{f-simu1}
\caption{Utility Division under Market Equilibrium.
Left: one PU $\{m\}$ and two SUs $\{n_1, n_2\}$. PU $m$ is matched to SU $n_1$ under the market equilibrium.
Right: two PUs $\{m_1,m_2\}$ and one SU $\{n\}$.  SU $n$ is matched to PU $m_1$ under the  market equilibrium.
The equilibrium utility division among the matched PU and SU is illustrated by the black bold line.}\label{f-simu1}
\vspace{-3mm}
 \end{figure*}

We first illustrate the utility division (or equivalently, resource exchange) under market equilibria to provide~an illustrative impression on the market equilibrium.

\revv{Figure \ref{f-simu1} illustrates the utility transfer functions (UTF) and the associated equilibrium utility division in the network with (i) one PU $\{m\}$ and two SUs $\{n_1, n_2\}$ (left figure) and (ii) two PUs $\{m_1,m_2\}$ and one SU $\{n\}$ (right figure).
\rev{Recall that the UTF function represents how a PU's maximal utility changes with the utility of the matched SU.}
In the left figure, \revmm{we choose the symmetric topology for the PU and two SUs (with $G_m^2 = -110$dBm, $G_{n(m)}^2= G_{m,n}^2= G_{n,m}^2= -90$dBm, $n\in\{n_1,n_2\}$) and different energy costs for two SUs (with $C_{n_1} = 1$ and  $C_{n_2} = 2$).}
In this case, PU $m$ is matched to SU $n_1$ under the market equilibrium.
In the right figure,
\revmm{we choose different topologies for the SU and two PUs (with $G_{m_1}^2 = -110$dBm, $G_{m_2}^2 = -100$dBm, $G_{n(m)}^2= G_{m,n}^2= G_{n,m}^2= -90$dBm, $m\in\{m_1,m_2\}$).}
%\revmm{we choose  $G_{m_1}^2 = -110$dBm, $G_{m_1}^2 = -100$dBm}, and
In this case, SU $n$ is matched to PU $m_1$ under the  market equilibrium. In both figures, the equilibrium utility division among the matched PU and SU is illustrated by the black bold line (between the circle and the square).
More specifically, the circle denotes the utility division under the Pareto-optimal equilibrium for PUs (\POEQ), and the square denotes the utility division under the Robust equilibrium for PUs (\RBEQ).
More detailed explanations for these equilibria are given below.

(1) \emph{{\POEQ} in the left of Figure \ref{f-simu1}:}
The {\POEQ} can be achieved by offering the lowest acceptable utility (i.e., zero in this example) to SU $n_1$. Obviously, the PU achieves its maximum utility under   {\POEQ} among all equilibria.

(2) \emph{{\RBEQ} in the left of Figure \ref{f-simu1}:}
The {\RBEQ} can be achieved by offering the highest achievable utility (i.e., 0.5 in this example) to SU $n_1$.
Note that due to the competition among SUs, SU $n_1$ cannot requesting a utility higher than 0.5 from the PU (which will leave the PU a utility lower than 0.8), otherwise, SU $n_2$ can successfully pair with the PU by requesting a utility slightly higher than zero (which will leave the PU a utility of 0.8).
%the PU will never offer a utility higher than 0.5 to SU $n_1$ (which will leave itself a utility lower than 0.8), otherwise, it can achieve a larger utility by matching to SU $n_2$ and offering SU $n_2$ a zero utility (which will leave itself a utility of 0.8).
This implies that the PU is guaranteed to obtain a utility of 0.8  (i.e., that under \RBEQ).
%Therefore, the {\RBEQ} provides the PU a guaranteed utility of 0.8.

(3) \emph{{\POEQ} in the right of Figure \ref{f-simu1}:}
The {\POEQ} can be achieved when PU $m_1$ offers the lowest acceptable utility (i.e., 1.7 in this example) to SU $n$.
Note that due to the competition among PUs, PU $m_1$ cannot offer a utility lower than 1.7 to SU $n$, otherwise, PU $m_2$ can offer a utility of 1.7 to SU $n$ such that SU $n$ accepts the offer of PU $m_2$.
Similarly, PU $m_1$ achieves its maximum utility under the  {\POEQ} among all equilibria.\footnote{Note PU $m_2$ achieves a zero utility under the \POEQ.}

(4) \emph{{\RBEQ} in the right of Figure \ref{f-simu1}:}
The {\RBEQ} can be achieved when PU $m_1$ offers the highest achievable utility (i.e., 2.2 in this example) to SU $n$.
Note that PU $m_1$ will never offer a utility higher than 2.2 to SU $n$, otherwise, it will achieve a negative utility.
Therefore, the {\RBEQ} provides PU $m_1$ a guaranteed  utility of zero.

From Figure \ref{f-simu1}, we can also see that the PUs' and SUs' equilibrium utilities are greatly affected by the numbers of PUs and SUs. Specifically, when the number of PUs is smaller than that of SUs (left figure), the PU is likely to achieve a large utility under the market equilibrium, benefiting from the competition of SUs.
When the number of PUs is larger than that of SUs (right figure), the PUs are likely to achieve small utilities under the market equilibrium, suffering  from the competition of PUs. Later in Figure \ref{f-simu4} we will show this observation more explicitly.}

\begin{figure}[tt]
\centering
\includegraphics[scale=.42]{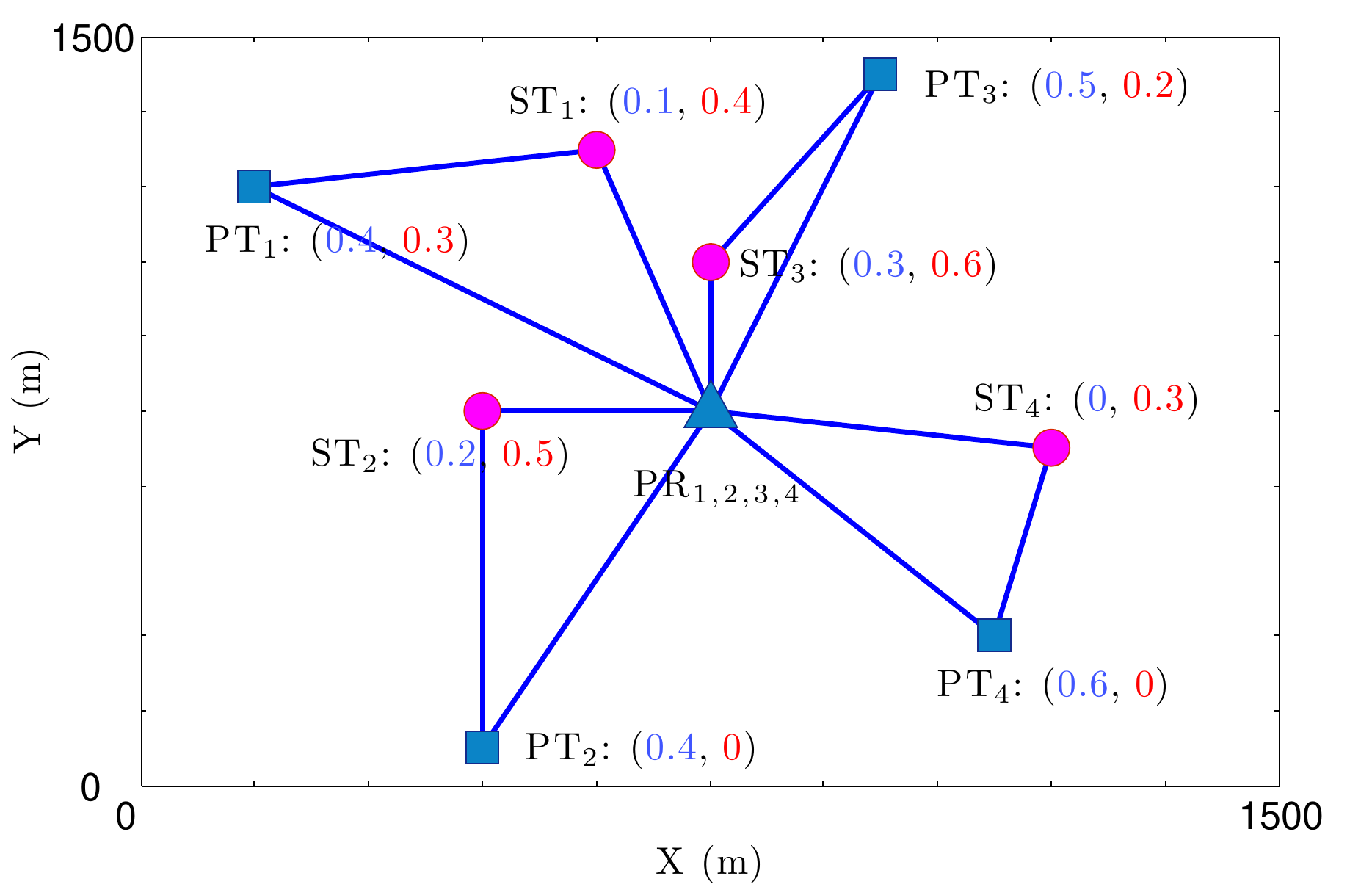}
    \caption{Relay Assignment under Market Equilibrium. Each PU $i$ is matched to SU $i$. The value associated with each PT$_i$ (or ST$_i$) denote PU $i$'s (or SU $i$'s) utilities under {\POEQ} (blue) and {\RBEQ} (red), respectively.} \label{f-relay}
    \vspace{-4mm}
\end{figure}

\subsubsection*{(2) Relay Assignment under Equilibrium}

Now we illustrate the relay assignment under the market equilibrium. To provide a clear impression, we simulate a network with 4 PUs and 4 SUs shown in Figure \ref{f-relay}, where primary receivers (PRs)  are co-located at the middle of the area (denoted by the trangle), and primary transmitters (PTs) are located close to the edges  of the area (denoted by squares).
\revmm{We use the default network setting except the channel gains, which are derived by the free-space path loss model in \cite{goldsmith2005wireless}.}
%, where all four PUs's receivers are co-located and denoted by PR$_{1,2,3,4}$.
%All other parameters are  same for all PUs (or SUs).

Figure \ref{f-relay} illustrates the network topology and the associated relay assignment (i.e., PU-SU matching) under the market equilibrium.
% achieved in different information scenarios.
In this example, SU $i$ is assigned to relay for PU $i$, $\forall i=1,2,3,4$, under the market equilibrium.
Note that PUs' and SUs' utilities are different under different market equilibria.
The values associated with each PT$_i$ (or ST$_i$) denote PU $i$'s (or SU $i$'s) utilities under the {\POEQ} (blue) and the {\RBEQ} (red), respectively. Specifically, under  the \POEQ, PUs' utilities are $\{0.4,0.4,0.5,0.6\}$, and SUs' utilities are $\{0.1,0.2,0.3,0\}$; Under the \RBEQ, PUs' utilities are $\{0.3,0,0.2,0\}$, and SUs' utilities are $\{0.4,0.5,0.6,0.3\}$.
Detailed discussions about such a utility difference are given below.

(1) \emph{PU 4 and SU 4:} PU 4 and SU 4 are far apart from other users, and  their interaction will be slightly affected by others' interactions.
Thus, under  the \POEQ, SU 4 will achieve the lowest acceptable utility (i.e., 0). Under the  \RBEQ, SU 4 will achieve the highest acceptable utility (i.e., 0.3), and   PU  4 can only achieve a zero utility.~~~~~~

(2) \emph{PUs 1 and 3:}  PU 1 is close to SUs \{1, 2, 3\}, and can potentially cooperate with these SUs.
Thus, SUs \{1, 2, 3\} will compete with each for PU 1.
Similar to the left figure of Figure \ref{f-simu1}, the PU can achieve a high utility (i.e., 0.3) under  the (worst-case) robust equilibrium  \RBEQ, benefiting from the competition of SUs.
Similarly, PU 3 can potentially cooperate with multiple SUs \{1, 3\}, and thus will also achieve a high utility (i.e., 0.2) under   the \RBEQ.

\subsection*{C. Implementation Issues}\label{sec:Implementation}

In this subsection, we briefly discuss the implementation issues of the proposed cooperative sharing scheme.

We first discuss how PUs obtain information in different information scenarios. As shown in Section \ref{sec:model:info}, we consider two information scenarios: complete and incomplete information.
First, complete information is an ideal benchmark (though not practical) case, where PUs know everything about the network.
Second, incomplete information is a more practical case, where  each PU knows only the local information, including its own channel gain, its relay channel gains (with neighboring SUs), and the neighboring SUs' channel gains.
The first class of local information can be easily obtained via measuring the average Received Signal Strength (RSS) from the PU's transmitter to   receiver.
The second class of local information can be obtained via   measuring  the average RSS from SUs's transmitters to the PU's transmitter and receiver, respectively.
The third class of local information (i.e., SUs' channel gains), however, cannot be measured by the PU directly.
Nevertheless, by measuring the average RSS from a SU's transmitter to the PU's transmitter and receiver jointly, the PU can estimate the SU transmitter's location (i.e., obtaining two candidate points); with the similar method, the PU can estimate the  SU receiver's location  (i.e., obtaining two candidate points).
Hence, there are at most four possibilities for an SU's topology.
 With certain additional knowledge (e.g., historical information or SU report),  the PU can derive the exact topology of the SU, and further calculate the SU's average channel gain.\footnote{\revmm{It is important to note that such an estimation is based on assumption that there is no deep fading or shadowing between the SU's transmitter and receiver, hence the channel gain is determined by the transmission distance only.
Otherwise, the PU may not be able to estimate the SU's own channel gain correctly. In that case, the estimation will be based on the SU's report directly.}}

We then discuss the time synchronization, which is a very important implementation issue.
In the existing literature of cooperative communications,  researchers have proposed many techniques to address this issue
(see, e.g.,  \cite{time-syn}).
In our model, we would like to clarify that we do not require all PUs and SUs to be fully synchronized during the whole interaction period.
\revww{As in many existing literature (e.g., [8]-[14]), we assume that there exists a common control channel for the necessary communications and interactions between PUs and SUs. Namely, PUs and SUs transmit all of the control signals (e.g., beacon, PU proposal, and SU response) on the common control channel using certain multiple access method such as CDMA, and continuously monitor the common control channel to decode the desired control signals.}
Hence, PUs and SUs can interact (on the common control channel) in the asynchronous manner during
the matching period (i.e., in the processing of Algorithms 2 and 3);
when a stable matching is reached (i.e., after the convergence of Algorithms 2 and 3), each pair of matched PU and SU then operate (on the PU's operating channel) in the synchronous manner during their cooperative transmission period (Phases I and II) and secondary transmission period (Phase III).
Such a synchronization (among a pair of matched PU and SU in Phases I, II, and III) can be achieved using different methods in
the existing cooperative communication literature (e.g., \cite{time-syn}).

\subsection*{D. Equilibrium Analysis for the Simplified Model}

Now we provide the complete equilibrium analysis for the simplified model. 

\subsubsection*{(1) Common \& Conflicting Interests on Equilibrium}

By Example \ref{examp:1}, we can find that there exists a \revh{unique} Pareto-optimal equilibrium for PUs: $\mu^a$, where \emph{every} PU achieves a no worse utility than in any other equilibrium.
Similarly, there exists a \revh{unique} Pareto-optimal equilibrium for SUs: $\mu^c $, where \emph{every} SU achieves a no worse utility than in any other equilibrium.
Lemma~\ref{lemma:opt-toy} shows that this observation is generally applicable.

\begin{lemma}[Optimality]\label{lemma:opt-toy} There always exists a unique Pareto-optimal equilibrium for PUs,
% (called {\POEQ}),
where {every} PU \revj{achieves its maximum utility among all equilibria}.
Similarly, there always exists a unique Pareto-optimal equilibrium for SUs, where {every} SU \revj{achieves its maximum utility among all equilibria.}
\end{lemma}

%Lemma \ref{lemma:opt-toy} can be proved by the Lattice Theory proposed by Kunth \cite{Knuth1976Marriage}. \rev{We skip the details due to space limit.}
%which states that if $\mu_1 $ and $\mu_2 $ are equilibria, then a new correspondence $\mu\triangleq \mu_1 \vee \mu_2$, where each PU (or each SU) points to the preferred partner in $\mu_1$ and $\mu_2$ (i.e., $\mu(m) = \mu_1(m)$ if PU $m$ prefers $\mu_1(m)$ more than $\mu_2(m)$, and $\mu(m) = \mu_2(m)$ otherwise), is not only a matching, but also an equilibrium (stable matching).
%%\footnotesc{The term ``matching'' implies an one-to-one correspondence  between PUs and SUs, that is, \emph{no} two PUs point to the same SU in $\mu$.}
%By repeatedly applying the Lattice Theory on any two equilibria, we finally address the PU-optimal (or SU-optimal) equilibrium.

%Recall Example \ref{examp:1}, we have $\mu^b \vee \mu^c = \mu^b$, $\mu^a \vee \mu^c = \mu^a$, and $\mu^a \vee \mu^b = \mu^a$, while $\mu^b \wedge \mu^c = \mu^c$, $\mu^a \wedge \mu^c = \mu^c$, and $\mu^a \wedge \mu^b = \mu^b$.

The following lemma further shows that among any two equilibria, the equilibrium that is better for all users on one side is always worse for all users on the other side.
%\rev{We can further show the conflicts between PUs and SUs.}
%
%We can further see from above example that all PUs like $\mu^a$ most and $\mu^c$ least, while all SUs like $\mu^a$ least and $\mu^c$ most. The following lemma generalizes this phenomenon.
\begin{lemma}[Conflicting Interest]\label{lemma:oppo-toy} \rev{For any   equilibria $\mu $ and $\mu^{\prime} $, \textit{all} PUs prefer $\mu $ \revj{to} $\mu^{\prime} $, if and only if \textit{all} SUs prefer $\mu^{\prime} $ \revj{to} $\mu $.}
\end{lemma}
%That is, any equilibrium that is better for \emph{all} PUs \rev{must be} worse for \emph{all} SUs, and vice versa.
%\com{I suggest removing the following sentence about proof, unless the above two results are our own. If the results are already in the literature, then we simply refer to the literature (and the readers will find the proofs from that literature).}

%\rev{Both Lemmas can be proved using the lattice theory in \cite{Knuth1976Marriage}.}
%\com{Lingjie: does this lemma includes the case where PUs and SUs have the common interest: PU1 likes SU1 most, PU2-SU2, PU3-SU3; SU1 likes PU1 most, SU2-PU2, SU3-PU3. Seems ``no less than'' has captured that. We may need to mention a sentence like ``This lemma'' includes common interest...}
%
Lemmas \ref{lemma:opt-toy} and \ref{lemma:oppo-toy} show that users on the same side
have a \emph{common interest} and users on the opposite sides have \emph{conflicting interests}, regarding the set of equilibria.\footnote{This observation was first discovered by Gale and Shapley in \cite{gale1962college}. The detailed proof  can also be referred to \cite{gale1962college}.}
%, since they all prefer the same optimal equilibrium, and (ii)  users on opposite sides have \emph{conflicting interests}, since an equilibrium that is better for one side is always worse for the other side.

%\subsection{Complete Information}
\subsubsection*{(2) Equilibrium under Complete Information}

%\footnotesc{Note that Gale and Shapley have proved the existence and optimality of stable matching (called equilibrium in our work) in a two-sided model without sidepay (called utility transferring in our work) under complete information (see Theorems 1 and 2 in \cite{gale1962college}). Here, we brief review their results as an illustration.}

\rev{Now we study which equilibrium will actually emerge (in the PU-proposal market) under  complete information.
In this case, any SU's misrepresentation (of its preference list) is not allowed, as PUs know the whole network information and thus know the complete preference lists of SUs.}

\revh{We first introduce the ``\emph{deferred acceptance}''   (DAC) algorithm  in \cite{gale1962college}, and then show
in Lemma \ref{lemma:converges} the DAC algorithm converges to the Pareto-optimal equilibrium for PUs ({\POEQ}).}

%that the DAC algorithm converges to the Pareto-optimal equilibrium for PUs ({\POEQ}).}

%with PUs proposing and SUs responding, and then show that DAC converges to an equilibrium and the resulting equilibrium is the only outcome under complete information
%\begin{algorithm}[Deferred Acceptance -- DAC]\label{algo:DAC}
%The deferred acceptance algorithm (DAC) works in an iterative fashion, where in every round
%
%(i) each PU proposes to its most preferred SU among those who have \textbf{not yet} rejected him; and
%
%(ii) each SU who receives multiple proposals rejects all but its most preferred.
%
%The algorithm stops when there is no PU being rejected.
%\end{algorithm}

\begin{algorithm}[h]
\small
\begin{algorithmic}[1]
\State \textbf{while} (\emph{at least one PU is rejected in previous round}) \textbf{do} \{
\State \hspace{2mm} Each PU \textbf{proposes} to the first SU in its preference list;
\State \hspace{2mm} Each SU \textbf{rejects} all proposals \revh{but} the most preferred;
\State \hspace{2mm} Each PU \textbf{updates} the preference list by removing the SU who just rejected him;~~~~\}
%\State \textbf{Converging Stage:}
%\State \hspace{2mm} No PU is rejected.
\end{algorithmic}
\caption{DAC}\label{algo:DAC} %\cite{gale1962college}
\end{algorithm}

\begin{lemma}\label{lemma:converges}
\rev{The DAC algorithm converges the Pareto-optimal equilibrium for PUs, i.e., {\POEQ}}.%, where {every} PU likes it at least as well as {any} other equilibria.
\end{lemma}

%By Lemma \ref{lemma:converges}, we impliedly show that the {\POEQ} is the only equilibrium that will emerge under complete information, since all PUs will follow the proposing rule in DAC and all SUs are \emph{obliged} to be truthful.
%\revj{We refer to the PU-optimal equilibrium achieved by DAC as \emph{{\POEQ}}.}

\revv{Notice that every PU can achieve a better utility under the {\POEQ} than under any other equilibrium.
That is, there does not exist another equilibrium, in which at least one PU achieves a higher utility than in the \POEQ.
Thus, all PUs are willing to follow the proposing rule in the DAC algorithm in order to achieve the {\POEQ}.
This implies that the {\POEQ} is the only equilibrium that will emerge under complete information.}

\subsubsection*{(3) Equilibrium under Incomplete Information}

Now let us consider the incomplete information scenario, where PUs know their local network information only, and thus cannot know the complete preference lists of SUs.
%We will consider the weakly and strongly incomplete information scenarios together in this illustrative example, since under both scenarios,
%In this case, each user can derive its own preference list (by (\ref{eq:PU_ut}) or (\ref{eq:SU_ut})) but {not} those of others.
%  (see Figure \ref{fig:information} for details).
%\footnote{Note that for the general model in Section~\ref{sec:main}, the analysis for these two scenarios will be different.}
Therefore, it is possible for SUs to misrepresent their preference lists to seek more utilities.
We show this by the following example.
%That is, two scenarios are equivalent to with regard to the users' preferences.

%The following example shows that the DAC algorithm cannot guarantee SUs' truthful information disclosure under incomplete information.
%Recall Example \ref{examp:1}. Here each user does not know others' preference lists.
%\rev{If SU $n_1$ misrepresents its preferences by $Q(n_1) = \{m_2, m_3\}$, DAC leads to the equilibrium $\mu^b$.
%Furthermore, if SU $n_1$ misrepresents its preferences by $Q(n_1) = \{m_2\}$, DAC leads to the SU-optimal equilibrium $\mu^c$. In both cases, SU $n_1$ achieves a higher utility from misrepresenting than from truth-telling.}
%In fact, in DAC algorithm,  SUs \emph{always} have incentives to misrepresent their preferences, since truth-telling will always yield the PU-optimal equilibrium which is the worst equilibrium for SUs (Lemma \ref{lemma:oppo-toy}).
\begin{example}\label{examp:2}
Consider the same model in Example \ref{examp:1}, except that here each PU does not know SUs' preference lists.
%The DAC algorithm  is expected to yield the PU-optimal equilibrium $\mu^a$, if all SUs \textbf{truthfully} represent their preferences.
\rev{Suppose SU $n_1$ misrepresents its preferences by $Q(n_1) = \{m_2, m_3\}$. Then it is easy to check that the   DAC algorithm  will lead to the equilibrium $\mu^b$.
Furthermore, if SU $n_1$ misrepresents its preferences by $Q(n_1) = \{m_2\}$, then the   DAC algorithm  will lead to the Pareto-optimal equilibrium $\mu^c$  for SUs. In both misrepresentations, SU $n_1$ achieves a higher utility than truth-telling.}
%
%Similarly, in an SU-proposal market, the PUs has the incentive to misrepresent their preferences as
%$$
%P(m_1) = \{n_1\},\ P(m_2) = \{n_2\},\ P(m_3) = \{n_3\},
%$$
%which results in the PU-optimal Equilibrium.
\hfill
$\square$
\end{example}

\revv{By the above example, we can see that different misrepresentation behaviors of SUs may lead to different market equilibria.
Unfortunately, characterizing all misrepresentation behaviors of SUs in the incomplete information scenario is an NP-hard problem. Hence, it is difficult to characterize which specific equilibrium will actually emerge in the incomplete information scenario, due to the uncertainty of SU misrepresentation.
To this end, we study the worst-case equilibrium for PUs, and  characterize a \emph{robust} equilibrium  for PUs ({\RBEQ}), which gives every PU a guaranteed utility under any possible misrepresentations of SUs.

Specifically, the robust equilibrium  {\RBEQ} can be achieved by a ``\emph{reversed deferred acceptance}'' (RDAC) algorithm as in  \cite{gale1962college}, which is same as the DAC algorithm except that it reverses the roles of PUs and SUs in proposing offers.
That is, in the RDAC algorithm, each SU proposes to its most preferred PU among those who have \emph{not yet} rejected him, and each PU accepts the most preferred proposal and rejects the others.  The details of the RDAC algorithm  are skipped due
to space limit.
By the symmetry between two algorithms, the RDAC algorithm converges to the Pareto-optimal equilibrium for SUs, which, by Lemma \ref{lemma:oppo-toy}, is the worst-case equilibrium for PUs.
Therefore, every PU is guaranteed to achieve a utility no worse than that in this worst-case equilibrium, under any possible misrepresentations of SUs.
In this sense, we refer to this worst-case equilibrium for PUs as the robust equilibrium for PUs.}

% if PUs \emph{commit} to truthfully represent their preferences.
%Thus, the RDAC algorithm (together with PUs' truthfulness commitments) is incentive compatible for SUs.\footnotesc{This is consistent with Roth's result in \cite{roth1984misrepresentation}, which shows that it is impossible to find a algorithm which provides \rev{incentives to all users simultaneously}. If one side is forced to be truthful (by commitment in our model), however, it is possible to find a algorithm (RDAC in our model) which provides \rev{incentives to users on the other side}.}
%\rev{As PUs are \revh{\emph{committed}} to be truthful in RDAC, we say that they act in a ``pessimistic'' manner (as this will lead to the SU-optimal equilibrium which is  the PU-worst equilibrium).} Thus, we refer to the equilibrium resulting from RDAC as \emph{PU-Pessimistic-EQ}.
%\revh{Obviously, PU-Pessimistic-EQ guarantees the worst-case utility of every PU under all possible misrepresentations of SUs.}
%We further notice that the equilibrium achieved in the RDAC guarantees every PU's achievable utility (under any possible equilibrium). That is,
%As every PU is guaranteed to achieve a utility no worse than that in this equilibrium (regardless of SUs' misrepresentation behaviors, we refer to this equilibrium as the robust equilibrium for PUs (denoted by {\RBEQ}).

%is a \emph{lower-bound} (in terms of PUs' utilities) of any equilibrium, since PU-Pessimistic-EQ is the SU-optimal equilibrium.}

\subsection*{E. Proof for Theorem \ref{lemma:necsuff1}}

\rev{We have already shown the necessity by the above analysis. Next we prove the sufficiency.}
% that the conditions in Theorem \ref{lemma:necsuff1} are necessary for equilibrium. Next we prove these conditions are also sufficient.
%The proof for necessity is directly from Lemma \ref{lemma:nece}. Now we provide the proof for sufficiency.

We first show that
%with above conditions,the matching would not be broken by any individual.
such a matching does not violate any PU or SU's \revj{IR condition}. This is because each SU $n$ gets a utility $\delta_n \geq \underline{\delta}_n \geq 0$ and each PU $m$ gets a utility $f_n^{\mu_n}(\delta_n) \geq  f_n^{\mu_n}(\overline{\delta}_n) \geq f_n^{\mu_n}(g_n^{\mu_n}(0))=0 $.
\revh{By (\ref{eq:opt_pu})}, \revv{we can further show that each pair of matched PU $\mu_n$ and SU $n$ have no incentive to change $\delta_n$, otherwise at least one user will lose certain utility (since an increase of one's utility will lead to a decrease of another's utility)}.

We then show that no PU or SU will deviate from the current matching by choosing a different partner. \revj{We prove this by contradiction.}
Suppose  that both PU $\mu_n$ (who is currently matched to SU $n$) and SU $k\neq n$ have incentives to break the matching by \revh{pairing up with} each other. Then there must \rev{exist a} value $\delta_k^{\prime} $ such that (i) $\delta_k^{\prime}  > \delta_k$ (i.e., SU $k$ achieves a larger utility with PU $\mu_n$) and (ii) $f_k^{\mu_n}(\delta_k^{\prime}) > f_n^{\mu_n}(\delta_n)$ (i.e., PU $\mu_n$ achieves a larger utility with SU $k$). \revh{On the other hand, we have
$$f_n^{\mu_n}(\delta_n) \geq  f_k^{\mu_n}(\delta_k) > f_k^{\mu_n}(\delta_k^{\prime}),$$
where the first \revj{inequality} follows directly from  $\delta_n \leq \overline{\delta}_n  $\com{why?}, and the second \revj{inequality} follows from the assumption $\delta_k^{\prime}  > \delta_k$. This leads to a contradiction.}

\subsection*{F. Proof for Lemma \ref{lemma:lattice}}

To prove $\{(\mu_n, \delta_n^X), \forall n \}$ is an equilibrium, we  only need to show that $\delta_n^X $ satisfies the conditions in Theorem \ref{lemma:necsuff1}.
By Theorem \ref{lemma:necsuff1}, we have: for all $ n\in \mathcal{N},$
$$
\underline{\delta}_n^i \leq \delta_n^i \leq \overline{\delta}_n^i, \quad   i=I,II.
$$
%as  $\{(\mu_n, \delta_n^I), \forall n \}$ and $\{(\mu_n, $ $ \delta_n^{II}), \forall n \}$ are equilibria.
%, where
%$\underline{\delta}_n^i$ and $\overline{\delta}_n^i$ are defined in Lemma \ref{lemma:nece}.

\revv{Without loss of generality, we consider an arbitrary matching pair, say, PU $\mu_n$ and SU $n$.
Suppose $ {\delta}_n^I\leq  {\delta}_n^{II}$}. Then $ {\delta}_n^X = \min( {\delta}_n^I,  {\delta}_n^{II}) = {\delta}_n^I$. On one hand,
\begin{equation*}
\begin{aligned}
\textstyle
 \delta_n^X = {\delta}_n^I \geq \underline{\delta}_n^I &  = \max_{m\neq \mu_n} \big\{g_n^{m}\big(f_{\mu_m}^m(\delta_{\mu_m}^I)\big), 0 \big\} \\
&  \geq \max_{m\neq \mu_n} \big\{g_n^{m}\big(f_{\mu_m}^m(\delta_{\mu_m}^X)\big), 0 \big\} \triangleq \underline{\delta}_n^X.
\end{aligned}
\end{equation*}
The  second line follows because $g_n^m\big(f_{\mu_m}^m(\cdot)\big)$ is an increasing function, and $\delta_{\mu_m}^X \leq \delta_{\mu_m}^I, \forall \mu_m$. On the other hand,
\begin{equation*}
\begin{aligned}
\textstyle
 \delta_n^X & = \min\big\{ {\delta}_n^I,  {\delta}_n^{II}\big\} \leq \min\big\{\overline{\delta}_n^I,  \overline{\delta}_n^{II}\big\} \\
  & = \min_{k\neq n}  \big\{{g_n^{\mu_n}} \big(f_k^{\mu_n}(\delta_k^I) \big),\ {g_n^{\mu_n}} \big(f_k^{\mu_n}(\delta_k^{II}) \big), {g_n^{\mu_n}}(0) \big\}
 \\
 &  = \min_{k\neq n}  \big\{{g_n^{\mu_n}} \big(f_k^{\mu_n}(\min(\delta_k^I, \delta_k^{II})) \big), {g_n^{\mu_n}}(0) \big\} \triangleq \overline{\delta}_n^X.
\end{aligned}
\end{equation*}
The last line follows because $h(\cdot) \triangleq {g_n^{\mu_n}} \big(f_k^{\mu_n}(\cdot) \big)$ is an increasing function, thus $\min (h(x),h(y) ) = h(\min(x,y))$.

\subsection*{G. Proof for Theorem \ref{lemma:opt}}

The existence and uniqueness of PU-optimal equilibrium can be easily proved by \rev{iteratively} applying \revj{Lemma \ref{lemma:lattice}} on any two equilibria. Next we prove \revj{that} the PU-optimal equilibrium $\{(\mu_n, \delta_n^*), \forall n \}$ \rev{satisfies} the conditions: $\delta_n^* = \underline{\delta}_n^* , \forall n$. \rev{If not}, there \rev{must exist} \revj{some $n$} such that $\delta_n^* > \underline{\delta}_n^* $, and $\delta_n^*$ can be further reduced without affecting the IC and IR conditions for equilibrium. Obviously, this violates the optimality of $\{(\mu_n, \delta_n^*), \forall n \}$.

\end{document}